\documentclass[twocolumn,twocolappendix]{aastex701}

\newcommand\sersic{S\'ersic}
\newcommand\msun{$M_{\odot}$}
\newcommand\deltalblt{$\Delta L_B/L_T$}

\newcommand\gampen{GaMPEN}

\newcommand\grizy{\textit{grizy}}
\newcommand\uband{\textit{u}}
\newcommand\gb{\textit{g}}
\newcommand\rb{\textit{r}}
\newcommand\ib{\textit{i}}
\newcommand\zb{\textit{z}}

\shorttitle{Environmental Sculpting of Galaxy Structure at Fixed Stellar Mass}
\shortauthors{Igel \& Ghosh et al.}

\usepackage{soul}
\usepackage{pifont}
\usepackage{enumitem}
\usepackage{bm}
\usepackage{amsmath}
\usepackage{gensymb}
\usepackage{subfigure}
\usepackage{multirow}
\usepackage{array}
\newcolumntype{P}[1]{>{\centering\arraybackslash}p{#1}}
\newcolumntype{M}[1]{>{\centering\arraybackslash}m{#1}}
\usepackage{hhline}
\usepackage{xcolor}
\usepackage{verbatim}
\PassOptionsToPackage{hyphens}{url}
\usepackage{hyperref}

\usepackage{xcolor} % to access the named colour LightGray
\definecolor{LightGray}{gray}{0.95}

\usepackage{CJK}

\font\bngxi=bang10 scaled 1100

\def\*#1*#2{o\null{#2}{#1}}

\def\d#1{\oalign{\smash{#1}\crcr\hidewidth{$\!$\rm.}\hidewidth}}

\def\sh#1{\setbox0=\hbox{#1}%
     \kern-.02em\copy0\kern-\wd0
     \kern.04em\copy0\kern-\wd0
     \kern-.02em\raise.0433em\box0 }

\begin{document}
\begin{CJK*}{UTF8}{bsmi}

\title{Environmental Sculpting of Galaxy Structure at Fixed Stellar Mass: A Multi-Scale Analysis Across Cosmic Time using 3 Million HSC Galaxies}
%%%%%ALTERNATIVE TITLES BELOW
%\title{Does Environment Impact Galaxy Structure at a Fixed Stellar Mass?: It's Complicated -- A Quantitative View using 3 Million HSC Galaxies}
%\title{The Morphology Environment Correlation at Fixed Stellar Mass: A Multi-Scale Analysis Across Cosmic Time using 3 Million HSC Galaxies}
%\title{The Evolving Role of Environment in Galaxy Structural Transformation: A Multi-Scale Analysis using 3 Million HSC Galaxies} 

\correspondingauthor{A. Ghosh: LSST-DA Catalyst Fellow}

\author[0009-0000-6412-3796, gname=Caitlin, sname=Igel]{Caitlin Igel (曹勝)}
\altaffiliation{C. Igel and A. Ghosh contributed equally to this work}
\affiliation{Department of Astronomy, University of Washington, Seattle, WA, USA}
\email{caitlin.igel@gmail.com}

\author[0000-0002-2525-9647, gname=Aritra, sname=Ghosh]{Aritra Ghosh ({\bngxi Airt/r \*gh*eaSh})}
\altaffiliation{C. Igel and A. Ghosh contributed equally to this work}
\affil{Department of Astronomy \& DiRAC Institute, University of Washington, Seattle, WA, USA}
\affil{eScience Institute, University of Washington, Seattle, WA, USA}
\email{aritrag@uw.edu}

\author[0000-0001-5576-8189]{Andrew J. Connolly}
\affiliation{Department of Astronomy \& DiRAC Institute, University of Washington, Seattle, WA, USA}
\affiliation{eScience Institute, University of Washington, Seattle, WA, USA}
\email{ajc@astro.washington.edu}

\author[0000-0002-4271-0364]{Brant Robertson}
\affil{Department of Astronomy and Astrophysics, University of California, Santa Cruz, CA, USA}
\email{brant@ucsc.edu }

\author[0000-0002-0745-9792]{C. Megan Urry}
\affil{Yale Center for Astronomy and Astrophysics, New Haven, CT, USA}
\affil{Department of Physics, Yale University, New Haven, CT, USA}
\email{meg.urry@yale.edu}

\author[0000-0002-9135-997X]{Louise O. V. Edwards}
\affil{Department of Physics, California Polytechnic State University, San Luis Obispo, CA, USA}
\email{ledwar04@calpoly.edu}

\author[0000-0003-4442-2750]{Rhythm Shimakawa}
\affil{Waseda Institute for Advanced Study (WIAS), Waseda University, Shinjuku, Tokyo, Japan}
\email{rhythm.shimakawa@aoni.waseda.jp }

\begin{abstract}
The extent to which galaxy structure is shaped by environment beyond the local universe, once stellar mass is controlled, remains an open question in galaxy evolution. We address this challenge using an unprecedentedly large sample of $\sim3$ million galaxies from the Hyper Suprime-Cam Subaru Strategic Program spanning $0.3 \leq z < 0.7$ with $\log \mathrm{M}/\mathrm{M}_{\odot} \geq 8.9$.
We correlate a mass-independent bulge-to-total ratio statistic with large-scale overdensity maps and cluster catalogs, propagating structural parameter posteriors through a Monte Carlo framework to robustly assess significance.
We confirm with $>5\sigma$ confidence that galaxy structure depends on environment at fixed stellar mass, but this dependence is secondary to stellar mass and varies with redshift, mass, and environmental scale. At $z < 0.5$, we detect no significant structural correlation with large-scale overdensity, but cluster galaxies show statistically significant bulge enhancement compared to mass-matched field galaxies, indicating cluster-specific processes such as ram-pressure stripping and cumulative tidal interactions dominate structural transformation at these epochs. At $z\geq0.5$, massive galaxies exhibit bulge-enhancement across both cluster- and large-scale environments, while lower-mass systems show enhancement only in cluster environments. This indicates that environmental mechanisms operate across broader spatial scales at earlier cosmic epochs, and enhanced merger rates, group preprocessing, and cosmic web stripping augment cluster-specific processes. By separating into star-forming and quiescent subsamples, we find nearly flat trends within each, demonstrating that the observed environmental effects arise from coupled morphological and star formation transformations. These results collectively reveal the multi-scale, epoch-dependent nature of environmental effects on galaxy structure.
\end{abstract}

\keywords{Extragalactic astronomy (506) --- Galaxies (573) --- Galaxy structure (622) --- Galaxy environments (2029)}

\section{Introduction} \label{sec:intro}
The structural properties of galaxies -- such as bulge prominence, radius, disk fraction, and overall morphology -- offer key observational clues to the physical mechanisms that govern galaxy evolution. One of the earliest and most enduring empirical patterns in this context is the morphology-environment correlation: the observation that elliptical and lenticular galaxies preferentially inhabit dense environments, while spiral systems dominate in lower-density regions. This trend was first observed by \citet{dressler1980} and has since served as a cornerstone for interpreting the role of environment in shaping galaxy structure.

The morphology - environment correlation is part of a broader network of environmental dependencies. Over the past several decades, cosmic environment has been shown to correlate with a wide array of galaxy properties, including color, star-formation, radius, gas content, nuclear activity, and kinematics \citep[e.g.,][]{Kauffmann2004TheGalaxies, Blanton2005RelationshipSurvey, Bamford2009, Peng2010MASSFUNCTION, Wetzel2012GalaxyBimodality, Darvish2017CosmicSatellites, Cleland2020TheGalaxies, Ghosh2024, Mercado2025EffectsGalaxies, Mhatre2025ActiveNeighbors, Asali2025TheEnvironments}. These trends have been interpreted as signatures of environmental mechanisms modifying galaxy properties through several pathways \citep[e.g.,][]{Gunn1972OnEvolution, Wechsler2018TheHalos, Boselli2022RamEnvironments, Wu2024HowEnvironment}. For a comprehensive review on this topic (on nearby galaxies), we refer the interested reader to \citet{Blanton2009}.

%physical processes that directly affect gas reservoirs \citep[such as ram-pressure stripping and strangulation; e.g.,][]{Gunn1972OnEvolution,Boselli2022RamEnvironments}, varying merger rates across different environmental densities \citep[e.g.,][]{Fakhouri2010TheSimulations, Jian2012ENVIRONMENTALUNIVERSE}, and complex environmental dependencies in the galaxy-halo connection that link stellar mass assembly to dark matter halo properties \citep[e.g.,][]{Wechsler2018TheHalos, Wu2024HowEnvironment}. 

Since \citeauthor{dressler1980}'s seminal work in \citeyear{dressler1980}, subsequent research has strengthened the established link between galaxy morphology and environmental density.
%with a growing body of evidence demonstrating that a galaxy's structural properties strongly correlate with its surrounding environmental density. These studies have extended observations to intermediate and high redshifts while exploring the dependence of the correlation on various galaxy properties. 
Some representative studies from the local universe and beyond are summarized in Appendix \ref{sec:ap:lit_survey}. However, a significant question has emerged regarding the primary driver of this relationship: Does environment directly influence galaxy morphology, or is stellar mass the dominant factor? This distinction is critical because more massive galaxies preferentially reside in denser environments, and galaxy morphology is also strongly tied to stellar mass, creating a potential confounding effect that could mask the true nature of environmental influence. Therefore, to investigate any fundamental physical link between galaxy morphology and environment, it is essential to control for stellar mass. 

At lower redshifts, studies conducted in the early 2000s established a consistent picture using SDSS data. These studies found that when stellar mass is properly controlled, environmental dependence becomes a secondary effect, with some studies confirming the presence of a weak correlation \citep[e.g.,][]{Kauffmann2004TheGalaxies, vanderWel2008, Bamford2009}. However, the picture becomes considerably more complex beyond the local universe, where many studies lack large samples to robustly control for stellar mass, while those that do attempt such controls report conflicting evidence, as summarized in Appendix \ref{sec:ap:lit_survey}. For instance, \citet{Kawinwanichakij2017Effectsup/sup} find significant environmental effects only at $\log\mathrm{M}/\mathrm{M}_{\odot} \gtrsim 10.25$, while at a similar redshift range \citet{Tasca2009} find the effect to be prevalent only at $\log\mathrm{M}/\mathrm{M}_{\odot} \lesssim 10.6$ -- directly contradicting each other. Other studies show equally diverse results: \citet{Sazonova2020TheClusters} report correlations in only $50\%$ of their cluster sample, while \citet{Chan2021The1.4} detect a correlation exclusively in quiescent galaxy sub-samples within narrow stellar mass ranges.  

The conflicting results described above stem from several fundamental methodological limitations that have constrained previous investigations. Most critically, the detection of weak secondary correlations --- such as environmental effects after controlling for stellar mass dependence --- requires both large sample sizes as well as robust uncertainty estimates on galaxy structural parameter measurements. Many earlier studies often worked with smaller samples and more modest uncertainty characterization --- appropriate for the data and algorithms available then --- but these constraints made it difficult to perform robust correlation analyses or to fully control for key confounding factors. Additionally, smaller samples make it difficult to account for the considerable diversity inherent in dense environments (e.g., varying cluster properties and dynamical states) and to adequately represent rare but potentially influential populations, such as the most massive galaxies that may drive environmental trends. Systematic discrepancies between measurements from different surveys or instruments (e.g., when comparing cluster and field galaxy samples) further complicate the interpretation of environmental effects. These methodological challenges and heterogeneity in findings at higher redshifts underscore the necessity for large, uniform datasets with well-characterized uncertainties to definitively resolve the role of environment versus stellar mass in driving morphological evolution beyond the local universe.

Recent advances in probabilistic machine learning (ML) combined with modern wide-field imaging surveys have now made it possible to overcome these obstacles. Deep, high-resolution imaging from the Hyper Suprime-Cam Subaru Strategic Program \citep[HSC-SSP;][]{Aihara2018, Miyazaki2018}, when paired with Bayesian ML frameworks such as the Galaxy Morphology Posterior Estimation Network \citep[GaMPEN;][]{Ghosh2022}, now enables measurements of well-calibrated posterior distributions for galaxy structural parameters across hundreds of square degrees beyond the local universe.

In this work, we leverage the comprehensive structural parameter catalog of HSC galaxies obtained using \gampen{} \citep{Ghosh2023} and correlate these with independent environmental tracers: large-scale density fields measured over 10 co-moving Mpc (cMpc) apertures \citep{Shimakawa2021Subaru1} as well as rich galaxy clusters (2 cMpc scale) detected via the CAMIRA algorithm \citep{Oguri2018}. Together, these complementary environmental tracers provide a multi-scale perspective, enabling us to quantify the morphology-environment relationship across both the large-scale environment as well as individual clusters. Using bulge-to-total light ratio ($L_B/L_T$) as our quantitative morphological indicator, we investigate how galaxy structure varies with environmental density at a fixed stellar mass, and how this relationship evolves over cosmic time. 

From the \citet{Ghosh2023} catalog, we assemble a large, uniform sample of $\sim3$ million galaxies spanning $0.3 \leq z < 0.7$ to a magnitude limit of $m_{\mathrm{AB}}=23$. This represents a sample that is one to four orders of magnitude larger than most previous studies, while extending stellar mass completeness limits by at least $\sim0.75$ dex. By incorporating the full Bayesian posteriors of $L_B/L_T$ of our large sample within a Monte-Carlo analysis framework, we are able to propagate uncertainties directly into our correlation analysis. This approach provides a rigorous, quantitative test of whether morphology correlates with environment once stellar mass is controlled for, enabling us to move beyond qualitative classifications and to robustly assess the relative roles of mass and environment in shaping galaxy structure.

In \S\ref{sec:data}, we describe our sample selection, along with the structural parameters, stellar masses, and environmental density tracers used in this work. \S\ref{sec:results} presents our Monte-Carlo correlation analysis framework and outlines the results of the above-mentioned correlation analysis between $L_B/L_T$ and environment. \S\ref{sec:conclusion} discusses the physical implications of these results, and we end with a summary of primary findings and future directions in \S\ref{sec:summary}. This work uses AB magnitudes and Planck18 cosmology \citep[$H_0=67.7$ km/s/Mpc, $\Omega_{\mathrm{m}}=0.311$, $\Omega_\Lambda=0.689$,][]{Aghanim2020}. 

\section{Data} \label{sec:data}
We use data from the Hyper Suprime-Cam (HSC) Subaru Strategic Program Public Data Release 2 \citep[PDR2;][]{Aihara2019}, along with structural parameter measurements from the \citet{Ghosh2023} HSC structural parameter catalog. Our sample selection criteria are described in \S\,\ref{subsec:sample}, and the various galaxy/environmental properties used in this analysis are detailed in \S\,\ref{subsec:struc_par} - \ref{subsec:mass}.

\subsection{Sample Selection} \label{subsec:sample}

We begin our sample selection process with the $\sim8$ million galaxies in the \citet{Ghosh2023} HSC-PDR2 structural parameter catalog. This catalog contains most HSC-Wide galaxies at $z \leq 0.75$ with $m_{\mathrm{AB}}\leq23$. We only include sources with $0.3\leq z<0.7$ that have detections above the $5\sigma$ threshold (based on the \texttt{cmodel} measurement) across all HSC bands. Additionally, we exclude galaxies located near bright stars or those affected by significant imaging artifacts, such as cosmic ray hits or saturated pixels. These issues are identified using the \texttt{mask\_s18a\_bright\_objectcenter} and \texttt{cleanflags\_any} parameters from PDR2. We note that these flags may preferentially remove the brightest and most massive members of galaxy clusters (BCGs) from our sample. The statistical impact of this selection on our cluster morphology measurements is quantified in \S\,\ref{subsec:lblt_v_env}.

We use photometric redshifts from the HSC photo-z catalog \citep{Nishizawa2020}, which were derived using the Bayesian template-fitting code \texttt{Mizuki} \citep{Tanaka2015}. For HSC-Wide galaxies with $m_{AB}\leq23$, the biweight dispersion of $\Delta z$ is $\leq0.05$, where $\Delta z=(z_{photo}-z_{spec})/(1+z{spec})$. To ensure a robust sample with reliable redshift estimates, we only include galaxies for which the parameter \texttt{photoz\_risk\_best} is below $0.1$.
%this criterion indicates a low probability of the photometric redshift deviating beyond $z_{\mathrm{true}} \pm 0.15(1+z_{\mathrm{true}})$. 
Following the recommendations of \citet{Nishizawa2020}, we further exclude galaxies whose best-fit photometric redshift model yields $\chi_{\nu}^2 > 5$. Additionally, consistent with \citet{Shimakawa2021}, we remove galaxies with specific star formation rates (sSFRs) exceeding $1\,\mathrm{Gyr}^{-1}$ within the redshift range $0.4 \leq z \leq 0.5$ to minimize potential contamination from Lyman-break galaxies at $z \sim 3$. 
%For galaxies at higher redshifts ($z>0.5$), contamination from Lyman-break galaxies (including \ib-band dropouts) is negligible due to our multi-band detection criteria, which require a detection in the \gb-band.

We divide our sample into redshift bins of width 0.1, resulting in four bins in the range $0.3\leq z<0.7$. We chose this bin width because it is $\geq2\times$ the bi-weight dispersion of $\Delta z$. Utilizing finer redshift bins could be feasible for a spectroscopic sub-sample; however, it is not possible for the large sample analyzed in this work. To get estimates of environmental density for our sample of galaxies, we cross-match the positions of the above galaxies with the fine grid of points in HSC-Wide, at which \cite{Shimakawa2021Subaru1} measured environmental densities. As recommended by the authors, we allow a maximum positional error of $64\arcsec$ independently within each redshift slice. This step ensures full consistency between the two survey areas for morphology and density estimation. %Note that the \citet{Shimakawa2021Subaru1} catalog included sources with $m \leq 23$, required $>5\sigma$ detections across all HSC bands, and followed a data-cleaning procedure similar to the one outlined above. This ensures that the structural parameter and environmental measurement catalogs used in this study are consistent with each other. 

Our final sample consists of 2,894,661 galaxies: 487,752 galaxies at $0.3\leq z<0.4$; 703,328 at $0.4\leq z<0.5$; 649,301 at $0.5\leq z<0.6$; and 1,054,280 at $0.6\leq z<0.7$. 
%Out of the $7.8$ million galaxies in \citet{Ghosh2023}, $\sim26\%$ of galaxies are excluded because we only include galaxies with $0.3\leq z<0.7$, with a further $\sim30\%$ being excluded in the positional cross-match with \citet{Shimakawa2021Subaru1}. The latter is because the environmental density catalog contains measurements in five of the seven HSC-Wide fields,  while the \citet{Ghosh2023} catalog measures structural parameters in all seven fields. All the other cuts combined result in a $\sim8\%$ reduction from the \citet{Ghosh2023} sample. 

\subsection{$L_B/L_T$ Measurements} \label{subsec:struc_par}

The primary structural parameter used in this work is the bulge-to-total light ratio ($L_B/L_T$), defined as the fraction of a galaxy's flux contained in its bulge component, derived from a two-component (bulge + disk) decomposition. Galaxies with higher $L_B/L_T$ values are more bulge-dominated, while those with lower values are more disk-dominated.

For every galaxy in our sample, we obtain the full posterior probability distribution of $L_B/L_T$ from the \citet{Ghosh2023} morphological catalog. This catalog uses the Galaxy Morphology Posterior Estimation Network \citep[\gampen;][]{Ghosh2022} to estimate Bayesian posteriors for various structural parameters of galaxies. \gampen{} has been extensively tested on both simulated and real HSC data \citep{Ghosh2022,Ghosh2023} and has been shown to produce accurate estimates with robust uncertainties. For $m\leq23$ HSC-Wide galaxies at $0.3 \leq z <0.7$, the dispersion in GaMPEN's prediction error [(predicted-true)/true] is $\leq0.16$ for $L_B/L_T$. GaMPEN's predicted posterior distributions for $L_B/L_T$ (i.e., uncertainty quantification) are extremely well-calibrated, with $\lesssim 5\%$ deviation, and outperform uncertainty estimates from light-profile fitting codes by up to $\sim60\%$ \citep[Figure 21 of ][]{Ghosh2023}.

\begin{figure}[htbp]
    \centering
    \includegraphics[width=1\linewidth]{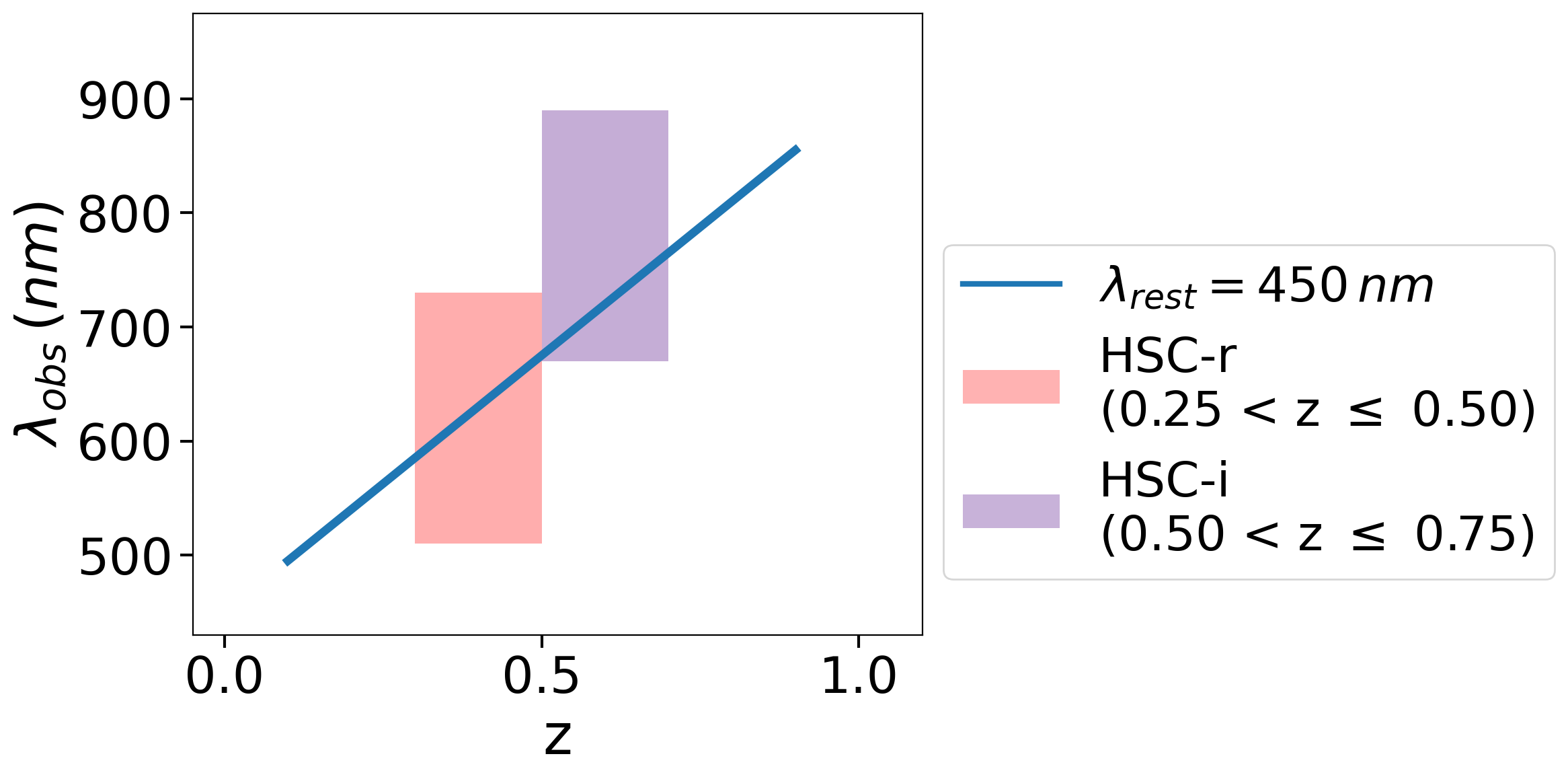}
    \caption{The filters used for morphological classification in \citet{Ghosh2023} at various redshifts are shown, along with the corresponding wavelength coverage of each filter. The blue line traces the observed wavelength corresponding to rest-frame $450\,\mathrm{nm}$ emission as a function of redshift. The selected filters enable consistent morphological analysis at a rest-frame wavelength of $\sim450\,\mathrm{nm}$ - corresponding to the rest-frame \gb{}-band - across the full redshift range of our sample ($0.3 \leq z < 0.7$). }
    \label{fig:graph1}
\end{figure}

To consistently use structural parameter measurements at a rest-frame wavelength of $\sim450\,nm$ (i.e., in the rest-frame \gb-band), we use measurements from the \rb-band for galaxies at $0.3\leq z < 0.5$; and from the \ib-band for galaxies at $0.5 \leq z < 0.7$, as illustrated in Figure \ref{fig:graph1}. 

\subsection{Density Measurements} \label{subsec:density}
To characterize the cosmic environment of our galaxy sample across a range of spatial scales, we employ two complementary density measurements:  a) large-scale projected overdensity estimates at scales of $10$\,cMpc from \citet{Shimakawa2021Subaru1}; and b) a catalog of red-sequence-selected galaxy clusters from \citet{Oguri2018}, which probe higher-density environments at smaller scales ($\sim2$\,cMpc).

%INTERNAL NOTE: FIGURE FROM new_paper_plots.ipynb
\begin{figure*}[htbp]
    \centering
    \includegraphics[width=0.87\linewidth]{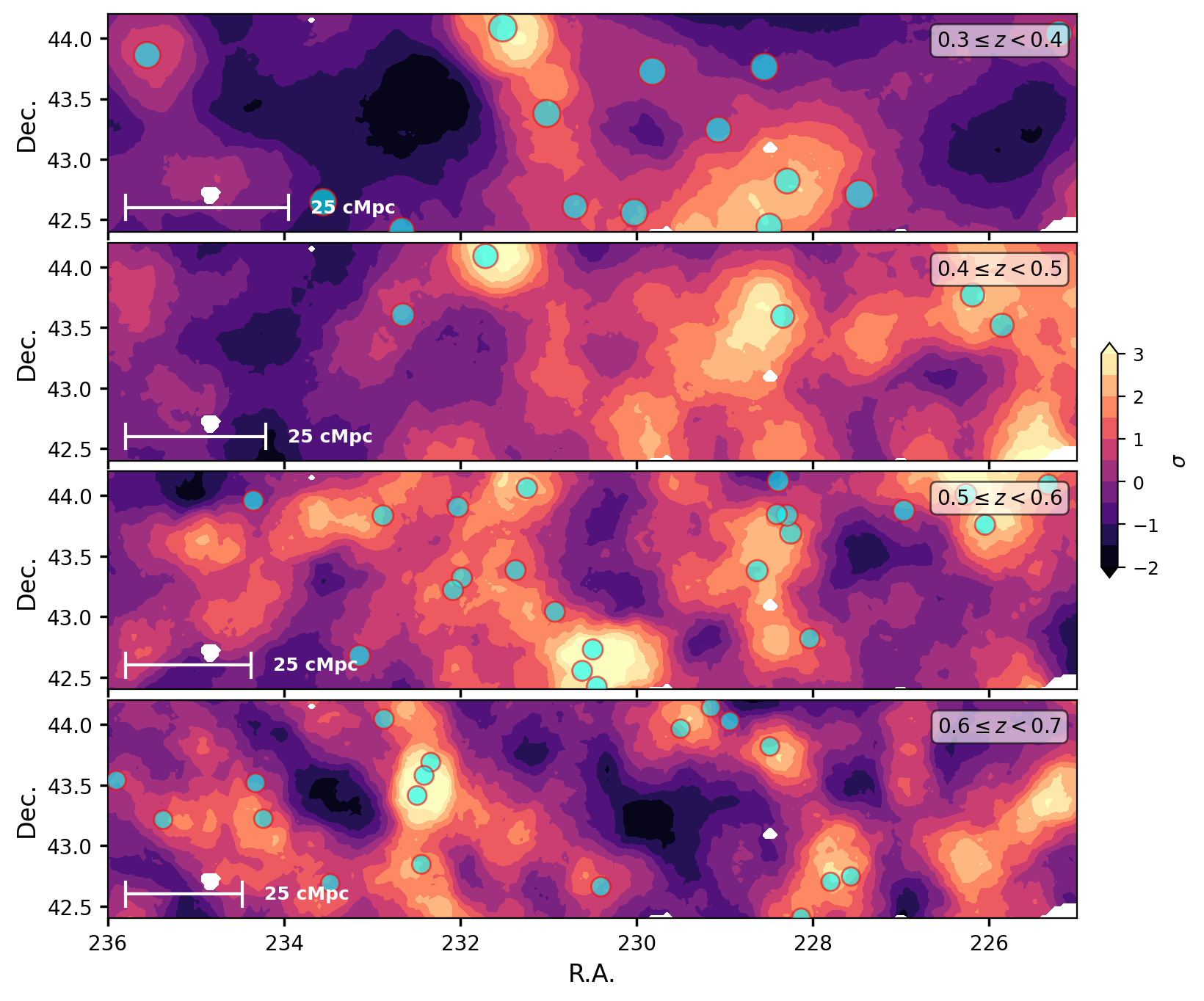}
    \caption{Projected two-dimensional overdensity maps for one of the five HSC-Wide regions analyzed in this study. Each row shows a distinct redshift slice, within which density excesses are computed. Colors indicate the density excess in standard deviation, measured within a fixed aperture of $r=10$ co-moving Mpc (cMpc). A 25\,cMpc reference scale is shown in each panel for context. White regions denote areas masked around bright stars, which were excluded from the density calculations; see \citet{Shimakawa2021Subaru1} for details. Cyan circles mark the positions of CAMIRA clusters in this field, each with a radius of 2\,cMpc - the distance used to associate galaxies in our sample with the clusters.}
    \label{fig:density_map}
\end{figure*}

We construct our large-scale density estimates using the projected two-dimensional overdensity maps from \citet{Shimakawa2021Subaru1}, which were measured over $\sim360$\,deg$^2$ of HSC Wide. Within each redshift slice, density measurements are provided on a uniform grid with a spatial resolution of $\sim1.5\arcmin$. At each grid point, the local environmental density is quantified using the standardized density contrast, $\sigma_{r=10\,{\rm cMpc}}$, computed within a top-hat aperture of radius $r = 10$\,cMpc. This quantity, listed as entry 128 in Table A1 of \citet{Shimakawa2021}, is defined as $(n_r - n_{\rm r, mean}) / \Sigma_r$, where $n_r$ is the galaxy number density within the aperture, and $n_{\rm r, mean}$ and $\Sigma_r$ are the mean and standard deviation of $n_r$ across the survey at the same redshift. We refer the reader to \S3.1 and Appendix A of \citet{Shimakawa2021} for further details on the construction and validation of the density field.

%Note that \citet{Shimakawa2021Subaru1} tested their method using mock galaxy catalogs, demonstrating that their projected overdensities reliably trace the total masses of embedded dark matter halos. They also validated their density field against mean lens shear signals from a weak lensing stacking analysis, thereby confirming the robustness of their environmental tracer.

The \citet{Oguri2018} cluster catalog was constructed using the Cluster-finding Algorithm based on Multi-band Identification of Red-sequence gAlaxies \citep[CAMIRA;][]{Oguri2014}. This algorithm identifies galaxy clusters by detecting overdensities of red-sequence galaxies and estimates their three-dimensional richness down to a limiting stellar mass of $10^{10.2}$\,\msun within $1\,h^{-1}$\,Mpc. In this work, we restrict our analysis to clusters with richness $\geq15$, which corresponds to a virial mass of $\gtrsim 10^{14}\,h^{-1}$\,\msun \citep{Oguri2018}. We adopt the cluster centers reported in the \citet{Oguri2018} catalog and associate galaxies in our sample that lie within a projected radius of $2$\,cMpc in the same redshift slice. This association radius is motivated by previous findings that the excess fraction of red galaxies extends out to $\lesssim2$\,cMpc from CAMIRA cluster centers \citep[see \S5.2 of ][]{Nishizawa2018}.

One of the five HSC-Wide fields analyzed in this study is shown as an example in Figure~\ref{fig:density_map}. The brightest regions in the map correspond to peaks in the projected density field, tracing some of the $\sim50$ large-scale overdensities (comparable to supercluster-embedded environments) identified by \citet{Shimakawa2021Subaru1}. The cyan circles indicate the locations of CAMIRA clusters within this field, with each circle representing a projected radius of 2\,cMpc. While most CAMIRA clusters coincide with the large-scale overdensity ridges, some deviations are expected due to the distinct nature of the two density tracers: The \citet{Shimakawa2021Subaru1} maps are sensitive to the smoothed distribution of galaxies over tens of megaparsecs, whereas the CAMIRA catalog identifies compact, high-richness systems based on red-sequence galaxy overdensities at cluster scales. Together, these two complementary tracers allow us to perform a comprehensive analysis on how galaxy structure responds to both large-scale overdensities in the cosmic-web and also compact cluster environments, capturing environmental effects across a broad range of scales.

\subsection{Stellar Masses} \label{subsec:mass}
We use stellar masses from the same Mizuki catalog that we obtained our redshift estimates from --- these are based on HSC \textit{grizy} photometry. We remind the reader that our study only includes sources with more than $5\sigma$ detection in all HSC bands, and we have also discarded sources with reduced chi-square $\chi_{\nu}^2 > 5$ for the best-fitting model.  \citet{Tanaka2018} and \citet{Shimakawa2021Subaru1} compared these Mizuki stellar masses to those from the NEWFIRM Medium-Band Survey \citep{Whitaker2011} and COSMOS2015 \citep{Laigle2016}, both based on $>30$-band photometry. They found Mizuki masses to be systematically higher at high redshift, though the offset is $\lesssim 0.1$ dex for $z \leq 0.7$, the redshift range of our study. We can attribute this discrepancy to template error functions and priors, as well as systematic differences in data. %Note that when comparing two independent surveys, stellar mass offsets of approximately $\sim0.2$–$0.3$ dex are common, even if both datasets feature deep photometry across multiple filters \citep[e.g.,][]{vanDokkum2014}.

\begin{figure}[h]
    \centering
    \includegraphics[width=1\linewidth]{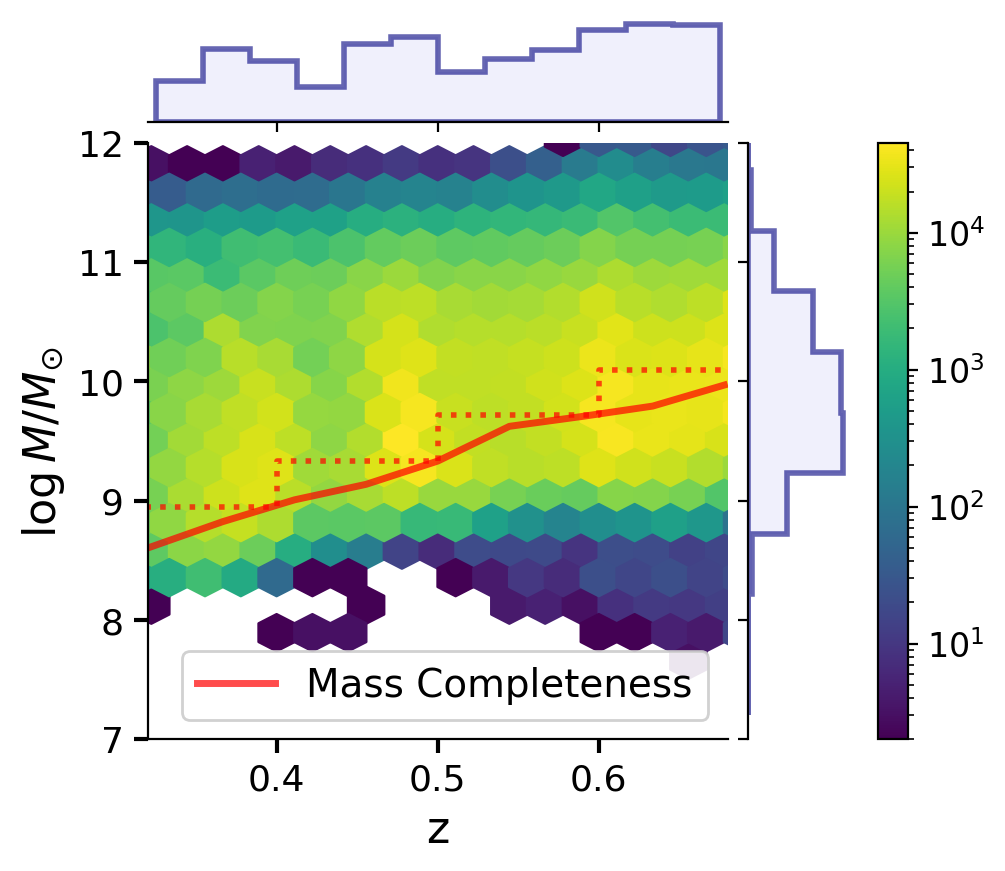}
    \caption{Distribution of galaxies in the stellar mass - redshift plane for our entire sample. The plane is divided into hexagonal bins of approximately equal area, with colors indicating the number of galaxies per bin as shown in the colorbar. Marginal histograms along each axis illustrate the one-dimensional distributions in stellar mass and redshift. The solid red curve shows the $90\%$ stellar mass completeness limit for our entire sample, while the dotted line denotes the lower limit ($M_c$) adopted for each redshift slice.}
    \label{fig:mass_completeness}
\end{figure}

We estimate the $90\%$ stellar mass completeness limit ($M_c$) of our sample following the methodology outlined in \cite{Pozzetti2010} and \cite{Weigel2016}. In narrow redshift bins, we compute the limiting stellar mass ($M_{\mathrm{lim}}$), defined as the stellar mass a galaxy would have if its apparent magnitude was $m=23$ (the adopted magnitude cut for our sample). We then select the faintest 20\% of galaxies in each bin and take $M_c$ to be the 90th percentile of $M_{\mathrm{lim}}$. The mass completeness limit of our entire sample is plotted (as the solid line) in Figure \ref{fig:mass_completeness}. Our sample is complete down to $\log (M/M_\odot)\sim8.5$ at $z=0.3$ and $\sim10$ at $z=0.7$. To ensure a conservative selection, we impose a step function above the derived completeness curve as shown in Figure \ref{fig:mass_completeness}, yielding final mass limits of $\log(M/M_\odot) = 8.95$, 9.33, 9.72, and 10.09 for the four redshift bins, respectively.

The marginal redshift histogram at the top of Figure~\ref{fig:mass_completeness} exhibits 
a few spikes. These are photometric redshift pileup artifacts arising from color-redshift degeneracies in the  template-fitting code. We have verified that they appear with consistent relative amplitude across all environments probed and therefore do not bias the morphology-environment correlations presented later.

\section{Results} \label{sec:results}
%The primary goal of this work is to robustly quantify the impact of environment on galaxy morphology beyond the local universe, while controlling for the primary confounding variable -- stellar mass -- using an unprecedentedly large sample. 
In \S \ref{sec:data}, we outlined how we assemble a dataset of $\sim3$ million HSC  galaxies with access to both their morphological and environmental properties.  Here, we present the trends and correlations that emerge from this analysis.

\subsection{Variation of $L_B/L_T$ with Stellar Mass} \label{subsec:lblt_v_mass}
\begin{figure*}[htbp]
    \centering
    \includegraphics[width=1\linewidth]{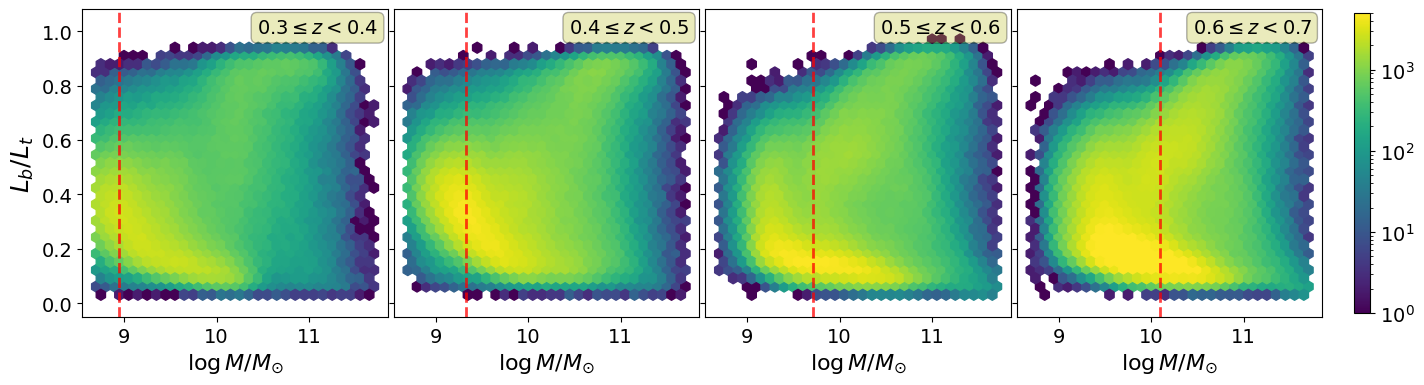}
    \caption{Distribution of bulge-to-total light ratio, $L_B/L_T$, as a function of stellar mass in the four different redshift slices. Colors indicate the number of galaxies per bin, shown on a logarithmic scale to reveal the full dynamic range of the distribution down to a single galaxy per bin. The red dashed vertical lines mark the stellar mass completeness limits for each redshift slice. Across all redshifts, the distribution is multi-modal with distinct populations of low-mass, disk-dominated galaxies; massive, bulge-dominated systems; and an intermediate sequence linking them, which is more prominent at higher redshifts.}
    \label{fig:lblt_m}
\end{figure*}

\begin{figure*}[htbp]
    \centering
    \includegraphics[width=1\linewidth]{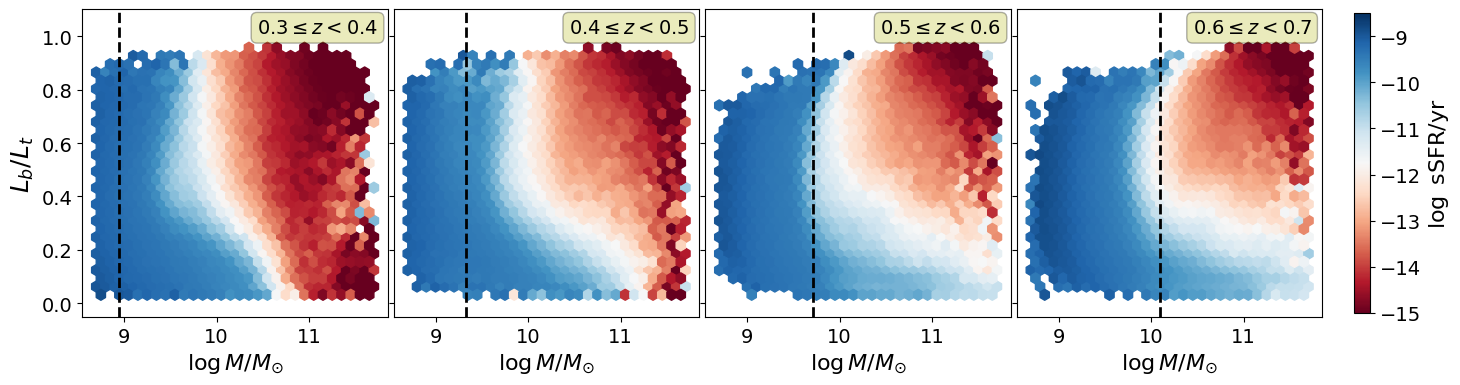}
    \caption{Specific star formation rate (sSFR) for galaxies in the $L_B/L_T$-stellar mass plane for the four redshift slices. Colors indicate the median sSFR in each hexagonal bin. The black dashed vertical lines mark the stellar mass completeness limits for each redshift slice. Low-mass, disk-dominated galaxies are uniformly star-forming, massive bulge-dominated systems are quenched, and the intermediate sequence shows a smooth transition, consistent with galaxies in various stages of transformation.}
    \label{fig:ssfr_lblt_mass}
\end{figure*}

In this work, we adopt the bulge-to-total light ratio, $L_B/L_T$, as our quantitative proxy for galaxy morphology. Unlike discrete, qualitative classifications (e.g., ``disk" vs. ``elliptical"), $L_B/L_T$ provides a continuous measure of structural dominance that captures the full diversity of galaxy morphologies. More importantly, because our measurements come from \gampen{}, we have access to well-calibrated probability distribution functions of $L_B/L_T$ for every galaxy in our sample. This enables us to incorporate measurement uncertainties directly into our correlation analysis, yielding statistically robust constraints that are not always possible when working with qualitative classifications or single-point estimates. 

Our overarching aim is to quantify the correlation between morphology and environment across a wide range of spatial scales: from large-scale overdensities in the cosmic web to compact, high-richness cluster environments. However, it is well established that more massive galaxies preferentially reside in overdense regions of the universe (see Appendix \ref{appendix:cdfs} for an analysis using our sample), and stellar mass is itself linked to the morphology of galaxies. Therefore, before interpreting any morphology-environment trend, we must first examine the direct relationship between $L_B/L_T$ and stellar mass. This step is essential for disentangling environmental effects from the impact of stellar mass on morphology. 

Figure \ref{fig:lblt_m} shows the distribution of $L_B/L_T$ as a function of stellar mass across the four redshift slices. At all redshifts, $L_B/L_T$ correlates strongly with stellar mass, but the distribution is multi-modal, with distinct concentrations rather than a smooth, monotonic trend. We see three distinct features in each panel: a) an L-shaped concentration of lower-mass disk-dominated galaxies in the lower left; b) a dense grouping of higher-mass bulge-dominated galaxies in the upper right; and c) a connecting sequence - more prominent at higher redshifts - bridging these two populations. This structure is similar to the long-recognized bimodality in galaxy color-mass space \citep[e.g.,][]{strateva_01,baldry_04,baldry_06,brammer_09}.

The state of star-formation in these structural populations is shown in Figure \ref{fig:ssfr_lblt_mass}, which shows the median specific star-formation rate (sSFR) from the Mizuki catalog in each bin of the $L_B/L_T$ - stellar-mass plane. To test the robustness of these sSFR measurements, we also constructed an alternate version of the figure using median quiescence scores, defined by a galaxy's distance from the star-forming/quiescent boundary in the rest-frame  $(u-r-z)_\mathrm{SDSS}$ color plane (see Appendix B and Figure 8 of \citealt{Ghosh2024}). Both methods reveal the same trends: the L-shaped concentration of disk-dominated galaxies is uniformly blue, indicative of ongoing active star formation; the dense cluster of massive, bulge-dominated systems is uniformly red, corresponding to quenched or nearly quenched systems; and the intermediate sequence shows a clear sSFR gradient from star-forming to quenched states, consistent with systems in various stages of morphological and star-formation transformation.

\begin{figure}[htbp]
    \centering
    \includegraphics[width=0.7\linewidth]{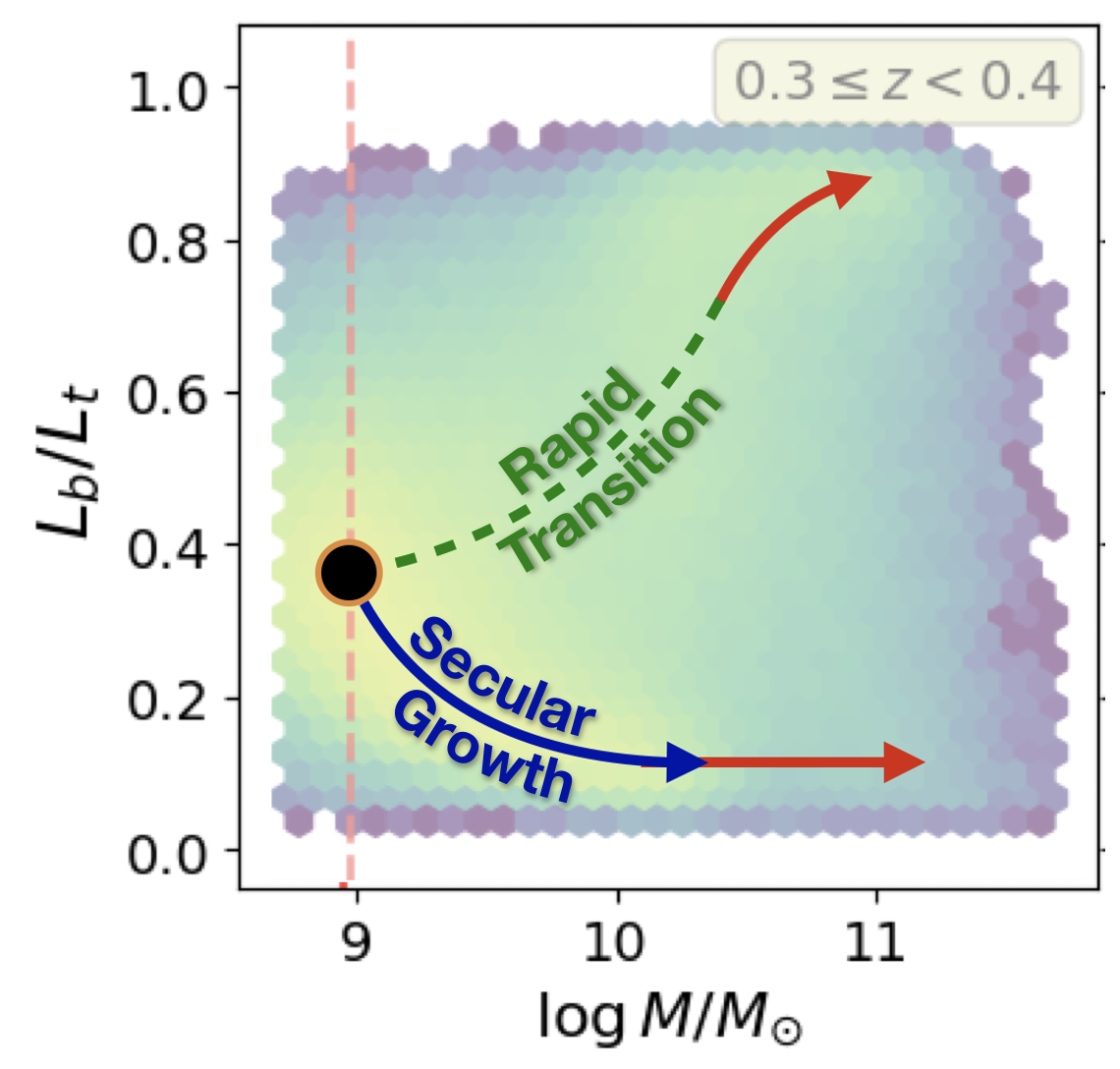}
    \caption{Schematic illustration of evolutionary pathways in the $L_B/L_T$-stellar mass plane (overlaid on the lowest-redshift slice). The blue arrow marks a secular growth track where disk-dominated galaxies build stellar mass while becoming more disk-dominated, with some quenching occurring at very high masses (for our low redshift sample). The green dashed arrow marks a rapid transformation track in which mergers or violent disk instabilities quench star formation and produce bulge-dominated systems. The greater prominence of the intermediate sequence at higher redshift suggests such transformations were more frequent or had shorter timescales in the past.}
    \label{fig:gal_evo_schematic}
\end{figure}

From a galaxy evolution perspective, the above-described features in Figures \ref{fig:lblt_m} and \ref{fig:ssfr_lblt_mass} reflect the interplay between stellar mass growth and morphological/structural transformation in galaxies, as schematically depicted in Figure \ref{fig:gal_evo_schematic}. We emphasize that Figure \ref{fig:gal_evo_schematic} is intended as a schematic, literature-motivated interpretation of the populations seen in Figures \ref{fig:lblt_m} and \ref{fig:ssfr_lblt_mass}, rather than as a direct evolutionary inference from our HSC data alone.

The L-shaped concentration in the lower left of each panel is dominated by disk-dominated, star-forming galaxies. Their strong prevalence across all redshift bins suggests that, at these epochs, these systems have avoided major morphological transformation. Rather, they evolve secularly---growing in stellar mass and becoming more disk-dominated. Note that at lower redshifts, we also see this population transitioning into a quenched state (bottom-right of the leftmost panel in Figure \ref{fig:ssfr_lblt_mass}). This may reflect galaxies whose fresh gas supply has been reduced, such that the remaining cold gas is gradually consumed and the stellar population reddens over multi-Gyr timescales \citep[e.g.,][]{schawinski_14}.

The cluster of massive, bulge-dominated, quiescent systems in the top-right of each panel is consistent with systems that have undergone non-secular rapid transformations---e.g., via major mergers, violent disk instabilities, or other gravitational processes---that not only altered the morphology of these galaxies but also led to the suppression of star formation in these systems. Previous work indicates that such rapid transformation channels can operate on timescales of $\lesssim 1$ Gyr \citep[e.g.,][]{Hopkins2008AEllipticals, Wild2009Post-starburstCuriosity}. The steep rise in $L_B/L_T$ at $\log \mathrm{M}/\mathrm{M}_{\odot} \gtrsim 10.5$ across all redshifts is consistent with the idea that morphological transformation becomes increasingly efficient above a characteristic mass scale.

The intermediate sequence connecting these two populations, more prominent at higher redshifts, represents galaxies in transitional phases on their way to becoming quenched and bulge-dominated. The greater visibility of this bridging population at higher redshifts indicates that the timescales for the processes driving these transitions (e.g., mergers) were shorter or occurred more frequently at earlier epochs, leading to a larger fraction of systems caught in the act of transformation.

Taken together, the multi-modal structure in Figure \ref{fig:lblt_m} points to the coexistence of at least two distinct evolutionary pathways for galaxies at these epochs. Note that this dual pathway for evolution is consistent with conclusions drawn from morphology-separated color-mass diagrams both in the local universe and beyond \citep[e.g.,][]{schawinski_14, powell_17, Ghosh2020}. 

In summary, the strong mass dependence of $L_B/L_T$ means that stellar mass must be controlled when examining morphology-environment trends. Without this, mass-driven structural correlations could be misattributed to environmental effects. In the sections that follow, we use mass-controlled samples to isolate the true influence of the environment.

\subsection{Variation of $L_B/L_T$ with Environment at a Fixed Stellar Mass} \label{subsec:lblt_v_env}
\begin{figure*}[htbp]
    \centering
    \includegraphics[width=1\linewidth]{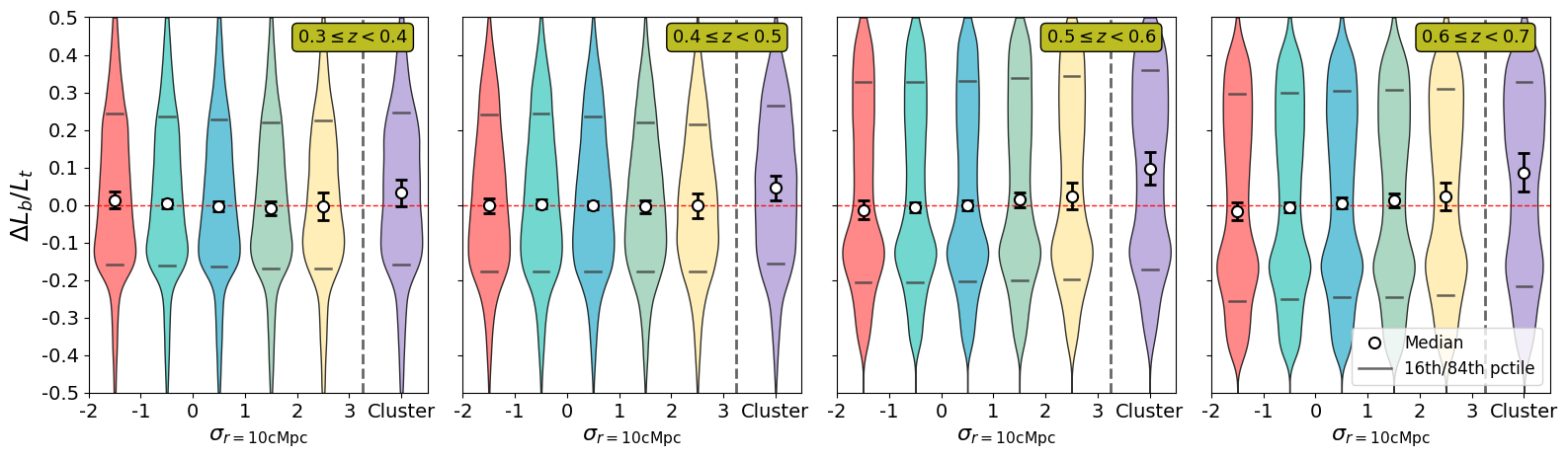}
    \caption{Distribution of $\Delta L_B/L_T$ as a function of environment for our full galaxy sample across the four redshift slices. The width of each violin traces the relative probability density of the distribution in one of six environment bins: five equally spaced bins spanning $\sigma_{r=10\,\mathrm{cMpc}} = [-2,\,3)$ and a ``Cluster'' bin containing galaxies within 2~cMpc of a CAMIRA cluster center. The horizontal bars demarcate the 16$^{\mathrm{th}}$--84$^{\mathrm{th}}$ percentile range of each distribution. The white points in each violin mark the median $\Delta L_B/L_T$, with error bars showing the $5\sigma$ uncertainty on the median estimate (scaled by a factor of 3 to remain visible at the scale of the plot). Overall, median $\Delta L_B/L_T$ shows little dependence on large-scale environment at $z < 0.5$, but consistently reveals bulge enhancement in cluster environments.}
    \label{fig:violin_all}
\end{figure*}

To characterize the impact of the environment on galaxy morphology, we define a new term

\begin{equation}
    \Delta L_B/L_T = L_B/L_T - \overline{L_B/L_T}(M,z),
\label{eq:delta_lblt}
\end{equation}

\noindent where $\Delta L_B/L_T$ quantifies the deviation in the galaxy's measured $L_B/L_T$ from the median value for all galaxies of similar stellar mass in the same redshift slice. By construction, this removes the mass dependence of $L_B/L_T$, enabling a cleaner test of whether environment drives additional variations beyond those set by stellar mass. To compute \deltalblt{}, we first determine the median bulge-to-total light ratio ($\overline{L_B/L_T}$) on a finely sampled stellar mass grid of 50 points within each redshift slice. At each grid point, we take all galaxies within $\log M/M_{\odot} \pm 0.125$ and calculate $\overline{L_B/L_T}$. This window-size is chosen to match the typical uncertainty in the Mizuki stellar mass estimates while maintaining sufficient numbers across the grid. Finally, each galaxy in our sample is assigned to the nearest grid point, and its \deltalblt{} is calculated following Eq. \ref{eq:delta_lblt}.

\begin{figure*}[htbp]
    \centering
    \includegraphics[width=1\linewidth]{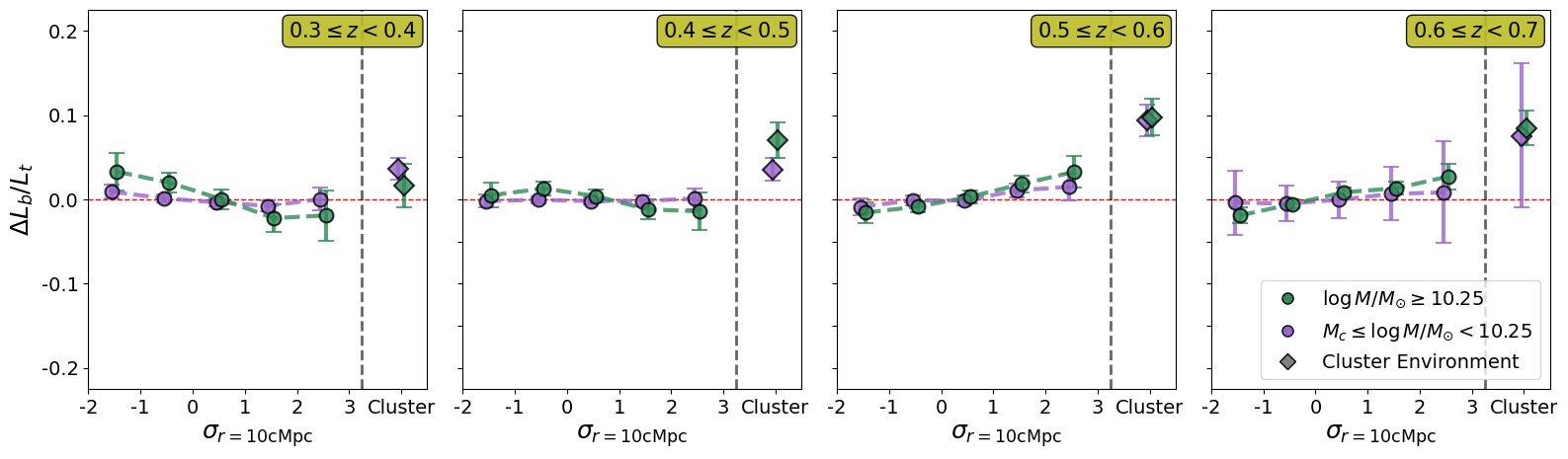}
    \caption{The median value of $\Delta L_B/L_T = L_B/L_T - \overline{L_B/L_T}(M)$ in each of the six environmental bins is shown for all redshift slices. 
    The galaxies are also separated into two mass bins, with the lower mass bin shown in purple and the higher mass bin in green.  
    Instead of the range of the distribution, the error bars on this plot denote the $5\sigma$ uncertainty on the median estimate. As a quantitative distillation of Figures \ref{fig:violin_all} and \ref{fig:violin_2mass_bins}, this plot shows that a monotonic environmental trend (across all environments) emerges only for massive galaxies at $z\geq0.5$, while clusters consistently exhibit elevated \deltalblt{} at almost all masses and redshifts.}
    \label{fig:deltla_lblt}
\end{figure*}

Figure \ref{fig:violin_all} presents the distribution of \deltalblt{} as a function of both large-scale and cluster environments. Within each redshift slice, galaxies are divided into six environmental bins. The first five bins are linearly spaced in projected overdensity within 10cMpc, $\sigma_{r=10\,{\rm cMpc}}$, as defined in \S\ref{subsec:density} and span the range $[-2,3)$. To incorporate the high-density regimes associated with massive, rich clusters within the same plot, we add a sixth bin -- shown as the rightmost point in each panel, labeled ``Cluster". This bin includes all galaxies within 2 cMpc of a CAMIRA cluster, as described in \S \ref{subsec:density}. While displayed along the same x-axis, the ``Cluster" point is not associated with a specific $\sigma_{r=10\,{\rm cMpc}}$ value, but rather serves as a qualitative flag. Taken together, the six points capture the full dynamic range of environments in our sample, from cosmic voids to dense cluster environments. Since $\Delta L_B/L_T$ for each galaxy is itself a posterior distribution (discussed further below), we use the mode (i.e., most probable value) of each distribution for calculating the overall $\Delta L_B/L_T$ distributions shown in Figure \ref{fig:violin_all}. 

To investigate whether the \deltalblt{} - environment correlation differs across different stellar-mass regimes, we divide our sample into two mass bins: $\mathrm{M}_{\rm c} \leq \log \mathrm{M}/\mathrm{M}_{\odot} < 10.25$, and $\log \mathrm{M}/\mathrm{M}_{\odot} \geq 10.25$. This chosen threshold ($\log \mathrm{M}\sim10.25\mathrm{M}_{\odot}$) corresponds to the 75th percentile of the overall stellar mass distribution. Figure \ref{fig:deltla_lblt} shows the mass-split median \deltalblt{} value for each environmental bin, along with the $5\sigma$ uncertainty on the estimate of the median, omitting the full violin shapes for clarity. The full violin distributions are shown in Figure \ref{fig:violin_2mass_bins} in Appendix \ref{appendix:mass_split_violins}.

Note that, throughout this paper, the ``$5\sigma$ uncertainty on the median'' denotes $5\times$ the half-width of the 16th--84th percentile credible interval on the median, estimated from the Monte Carlo draws. The relatively large error bars seen for the lower-mass bin in the highest redshift slice arise because the stellar-mass completeness limit in this bin, $\log\mathrm{M}_c = 10.09\,\mathrm{M}_\odot$, lies very close to the division between our two mass bins, leaving this specific mass bin substantially narrower and with fewer galaxies compared to the others.

For $z<0.5$, there is no clear monotonic variation in \deltalblt{} with environment among the large-scale environment bins (circular points). However, the cluster points exhibit a systematic positive offset from \deltalblt{}$=0$, indicating an excess of bulge-dominated systems in the richest environments at a fixed stellar mass. For $z\geq0.5$, the higher mass bin shows a monotonic trend in both redshift slices, rising from $\sigma_{r=10\,{\rm cMpc}} = -2$ to the ``Cluster" point. This suggests that, at these epochs and mass regimes, denser environments across a range of scales systematically host more bulge-dominated galaxies, even after controlling for stellar mass. In both higher-redshift slices, the ``Cluster" points alone show a statistically significant positive deviation from zero. Except in the lowest-redshift slice, the ``Cluster" medians are higher than those of any large-scale environment bin, implying that cluster-scale environments exert a stronger influence on galaxy morphology than large-scale density at the scale of the cosmic web. We examined the impact of the line-of-sight depth on the cluster results by repeating the analysis with a cylinder half-height set to the individual photometric redshift uncertainty of each galaxy, and found $\Delta L_B/L_T$ trends consistent with those from the fiducial selection across all redshift and stellar mass bins (mean absolute shift of $0.008$ in $\Delta L_B/L_T$).

\begin{figure}[htbp]
    \centering
    \includegraphics[width=1\linewidth]{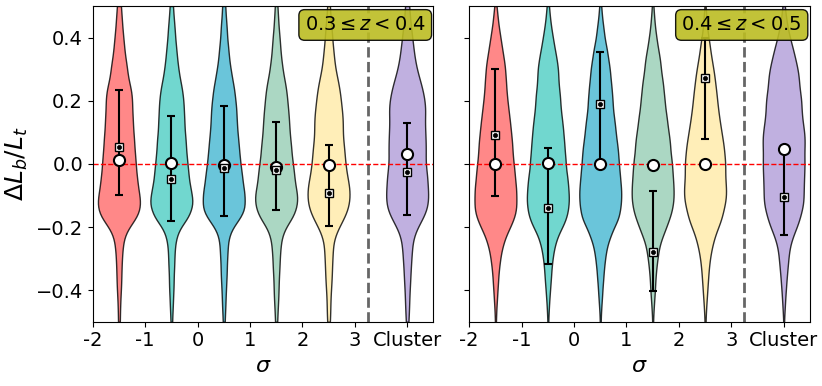}
    \caption{ Example demonstrating that each point contributing to the violins in Figure \ref{fig:violin_all}  represents the median of an underlying \deltalblt{} posterior distribution. In this example, for each violin we randomly select one galaxy and show the median of its \deltalblt{} distribution (grey squares) along with the $16^{\mathrm{th}}-84^{\mathrm{th}}$ percentile range. This arises from the fact that we have access to well-calibrated posterior distributions of $L_B/L_T$ from \gampen{} for every galaxy in our sample.}
    \label{fig:deltla_lblt_with_dist}
\end{figure}

  \begin{deluxetable*}{c|c|cccc}[htbp]                                                                                                                                
  \tablecolumns{6}                                                                                                                                                    
  \tablehead{Mass Range & Statistical & $0.3 \leq z < 0.4$ & $0.4 \leq z < 0.5$ & $0.5 \leq z < 0.6$ & $0.6 \leq z < 0.7$ \\
            ($\log M/M_{\odot}$) & Test Metrics & & & & }                                                                                                             
  \startdata                                                                                                                                                          
      \hline                                                                                                                                                          
      \hline                                                                                                                                                          
      \multirow[c]{3}{*}{$\left[M_{\rm c},10.25\right)$} & $\bar{z}_{\rm JT}$ & $-3.9\pm0.9$ & $+0.7\pm0.9$ & $+4.5\pm0.8$ & $+1.9\pm1.7$ \\                                                  
                                      & $p < 5\sigma_p$ & $11.1\%$ & $0\%$ & $33.5\%$ & $4.1\%$ \\                     
                                      & Confirmed? & & & $(\checkmark)$ & \\                                                                                                         
      \hline                                                                                                                                                          
      \multirow{3}{*}{$\geq10.25$} & $\bar{z}_{\rm JT}$ & $-4.9\pm0.8$ & $-1.0\pm0.8$ & $+9.0\pm0.8$ & $+9.8\pm0.8$ \\                                                                        
                                   & $p < 5\sigma_p$ & $45.0\%$ & $0\%$ & $100\%$ & $100\%$ \\                                                                           
                                   & Confirmed? & & & \checkmark & \checkmark \\                                                                                      
      \hline                                                                    
      \hline                                                                                                                                                          
  \enddata                                               
  \caption{Summary of Jonckheere--Terpstra test results for the relationship between \deltalblt{} and the six ordered environment bins. $\bar{z}_\mathrm{JT} \pm \sigma_z$ gives the median $\pm$ standard deviation of the JT $z$-statistic across 5000 Monte Carlo realizations; positive values indicate a positive monotonic trend of \deltalblt{}
   with environment, negative values indicate the opposite. $p<5\sigma_p$ lists the fraction of realizations with $p<5.74\times10^{-7}$, the $5\sigma$ significance   
  threshold. The Confirmed? row marks bins where ($\bar{z}_{\rm JT} - \sigma_z) \geq 5$ \textit{and} more than $99.99994\%$ of realizations have $p < 5\sigma_p$, indicating a
  statistically significant positive monotonic trend at the $5\sigma$ level or higher. Cases marked $(\checkmark)$ fail the $5\sigma$ criterion, but are confirmed under the less-conservative Benjamini--Hochberg FDR procedure.}                                                                           
  \label{tab:separman}                                   
  \end{deluxetable*}

While Figures \ref{fig:violin_all} and \ref{fig:deltla_lblt} provide a qualitative view of how \deltalblt{} varies with environment, our large sample also allows a statistically rigorous quantitative assessment. As illustrated in Figure \ref{fig:deltla_lblt_with_dist}, \deltalblt{} for each galaxy is itself a posterior distribution arising from \gampen{}'s full $L_B/L_T$ PDFs. To propagate these uncertainties, we draw 5000 samples from each galaxy's $L_B/L_T$ posterior and construct 5000 dataset realizations; within each we recompute $\overline{L_B/L_T}(M,z)$ and evaluate \deltalblt{} for every galaxy via Eq.~\ref{eq:delta_lblt}. To each realization we apply the Jonckheere--Terpstra (JT) test \citep{Jonckheere1954}, the standard nonparametric method for detecting a monotonic trend in a continuous response variable across ordered groups. The JT statistic $J$ counts the number of concordant pairs --- galaxy pairs where the member in the higher-ranked environment has higher \deltalblt{} --- and is standardized to $z_{\rm JT} = (J - \mathbb{E}[J])/\sqrt{\mathrm{Var}[J]}$, which follows $\mathcal{N}(0,1)$ under the null hypothesis of no trend. We order the five $\sigma_{r=10,\mathrm{cMpc}}$ bins from underdense to overdense and designate the cluster bin as the sixth and highest-ranked group, so a single test spans the full dynamic range of environments.                                           
           
The 5000 realizations yield a distribution of $z_{\rm JT}$ and $p$, summarized in Table~\ref{tab:separman}. $\bar{z}_{\rm JT}$ is the median standardized JT statistic; positive values indicate that \deltalblt{} tends to increase with environment rank, and negative values indicate the reverse. $p<5\sigma_p$ is the fraction of realizations with $p < 5.74\times10^{-7}$ (the $5\sigma$ threshold). A \checkmark{} in ``Confirmed?" signifies that we can reject the null hypothesis of no positive monotonic trend at a significance level of $\geq 5\sigma$; this requires both ($\bar{z}_{\rm JT}-\sigma_z) \geq 5$ and more than $99.99994\%$ of realizations satisfying $p < 5\sigma_p$.

\begin{deluxetable*}{c|c|cccc}[htbp]
\tablecolumns{6}
\tablehead{Mass Range & & $0.3 \leq z < 0.4$ & $0.4 \leq z < 0.5$ & $0.5 \leq z < 0.6$ & $0.6 \leq z < 0.7$ \\ 
          ($\log M/M_{\odot}$) & & & & & }
\startdata
    \hline
    \hline
                                                 & $\overline{W}$   & $\sim10^{7}$ & $\sim10^{7}$ & $\sim10^{6}$ & $\sim10^{6}$ \\
                                    Cluster Only & $(\overline{W}-W_0)/\sigma_W$ & $24.5$ & $25.0$ & $22.2$ & $9.8$  \\
                                    $\left[M_{\rm c},10.25\right)$ & $p < 5\sigma_p$ & $100\%$ & $100\%$& $100\%$ & $100\%$ \\
                                    %& $p < 3\sigma_p$ & $95.6\%$ & $0.2\%$& $98.9\%$ & $39.0\%$ \\
                                    & Confirmed? & \checkmark & \checkmark & \checkmark & \checkmark \\
    \hline
                                 & $\overline{W}$ & $\sim10^{6}$ & $\sim10^{7}$ & $\sim10^{7}$ & $\sim10^{7}$ \\
                                 Cluster Only & $(\overline{W}-W_0)/\sigma_W$ & $2.5$ & $10.6$ & $16.4$ & $16.8$  \\
                                 $\geq10.25$ & $p < 5\sigma_p$ & $0\%$ & $100\%$& $100\%$ & $100\%$ \\
                                 %& $p < 3\sigma_p$  & $99.3\%$ & $3.8\%$& $100\%$ & $100\%$ \\
                                 & Confirmed? &  & \checkmark & \checkmark & \checkmark \\
    \hline
    \hline
\enddata
\caption{Summary of Wilcoxon signed-rank test results for the cluster-environment bin. 
\(\bar{W}\) is the order-of-magnitude median value of the Wilcoxon statistic, and \((\overline{W} - W_{0})/\sigma_{W}\) gives the standardized displacement of \(\overline{W}\) from the expected null statistic $W_{0}$. The row \(p < 5\sigma_{p}\) lists the fraction of realizations with \(p < 5.74\times 10^{-7}\), the \(5\sigma\) significance threshold. 
The Confirmed? row marks bins where \(|(\overline{W} - W_{0})/\sigma_{W}| \geq 5\) \textit{and} more than \(99.99994\%\) of realizations have \(p < 5\sigma_{p}\), indicating a statistically significant deviation of the cluster median \(\Delta L_B/L_T\) from zero at the \(5\sigma\) level or higher.
}
\label{tab:wilcoxon}
\end{deluxetable*}

%For the cluster bin, we separately perform a different non-parametric test that quantifies whether \deltalblt{} for the cluster points significantly deviates from \deltalblt$=0$ and by how much. Physically, this addresses whether rich cluster environments drive measurable structural transformations on their own beyond those associated with large-scale overdensity alone. To test this, we restrict the sample to CAMIRA cluster members and apply the Wilcoxon signed-rank test \citep{wilcoxon}. The test orders galaxies by the size of their $|$\deltalblt{}$|$ offset from zero, assigns ranks, and then sums the ranks to produce the Wilcoxon statistic $W$. Under the null hypothesis of no offset, the expected statistic is $W_0=n(n+1)/4$, where $n$ is the number of galaxies in the sample. The Wilcoxon test also reports a $p$-value, which signifies the probability of obtaining a $W$ statistic at least this extreme under the null hypothesis that \deltalblt$=0$. Similar to the Spearman analysis, we propagate uncertainties by creating 5000 different realizations of the cluster bin and performing the Wilcoxon test for each realization. The resulting distributions of $W$
%and $p$ across these realizations are summarized in Table \ref{tab:wilcoxon}, using the same format as Table \ref{tab:separman} but with $\rho$ replaced by $W$ as the primary test statistic. With these two robust statistical analysis frameworks in place, we now examine the significance of the trends that emerge out of Figures \ref{fig:violin_all} - \ref{fig:deltla_lblt} and Tables \ref{tab:separman} - \ref{tab:wilcoxon}.}

For the cluster bin, we separately apply the Wilcoxon signed-rank test \citep{wilcoxon} to assess whether cluster galaxies' \deltalblt{} is significantly offset from zero -- addressing whether rich clusters drive structural transformation beyond what large-scale overdensity alone produces. The test orders galaxies by the size of their $|$\deltalblt{}$|$ offset from zero, assigns ranks, and then sums the ranks to produce the Wilcoxon statistic $W$. Under the null hypothesis of no offset, the expected statistic is $W_0=n(n+1)/4$, where $n$ is the number of galaxies in the sample. The Wilcoxon test also reports a $p$-value, which signifies the probability of obtaining a $W$ statistic at least this extreme under the null hypothesis that \deltalblt$=0$. As with the large-scale correlation analysis above, we propagate uncertainties by constructing 5000 dataset realizations and performing the test on each; the resulting distributions $W$ and $p$ are summarized in Table~\ref{tab:wilcoxon} using the same format as Table~\ref{tab:separman}, with $z_{\rm JT}$ replaced by $W$.

As a less conservative complement to the $5\sigma$ criterion, we also apply the Benjamini--Hochberg false discovery rate procedure \citep[BH-FDR;][]{Benjamini1995ControllingTesting} at $\alpha=0.05$ (the conventional significance level) across the full family of 16 simultaneous tests (four redshift bins $\times$ two mass bins $\times$ two test types). All $5\sigma$-confirmed results are also confirmed under BH-FDR; one additional detection is newly confirmed and is discussed below, marked with $(\checkmark)$ in Table~\ref{tab:separman}.    

For $z<0.5$, Table \ref{tab:separman} confirms that there is no statistically significant monotonic dependence of \deltalblt{} on environment across the six bins considered. In other words, when all environments---from low-density voids to rich clusters---are examined together, we do not find evidence for a smooth, continuous relationship between morphology and overdensity once stellar mass is controlled for. Table \ref{tab:wilcoxon}, however, demonstrates that when focusing specifically on rich cluster environments at $z < 0.5$, the signal becomes clear: in almost all cases, galaxies in clusters are more bulge-dominated than mass-matched field galaxies, with significance exceeding the five-sigma level. This result indicates that environmental structural transformations at these epochs are not generally linked to large-scale environmental density but rather are driven primarily by the specific physical processes active in rich clusters, such as ram-pressure stripping or repeated tidal interactions. A notable exception to the above is the highest-mass bin in the lowest redshift slice, where the Wilcoxon test does not indicate a significant positive offset. As seen in the cluster points of the higher-mass bin (lower panel, purple violins) of Figure \ref{fig:violin_2mass_bins}, the \deltalblt{} distribution peaks (i.e., is widest) at \deltalblt{} $\sim0.2$, whereas at higher redshifts, it peaks at $\sim0.3-0.4$. This lower mode value cannot overcome the long tail of negative \deltalblt{} values, suppressing the test statistic and yielding a non-significant result. Physically, this suppression of the cluster-induced offset might indicate that by $z\sim0.3-0.4$, the most massive galaxies are already largely bulge-dominated within all environments where they exist. In such a regime, the role of clusters is diminished, since the structural transformation that clusters accelerate at earlier epochs (at similar stellar masses) has already saturated, leaving little residual difference between massive cluster and field galaxies.

For $z\geq0.5$, the overall nature of the results changes. As suggested by the right-hand panels of Figure \ref{fig:deltla_lblt}, Table \ref{tab:separman} confirms with more than $5\sigma$ confidence that for $z\geq0.5$ and $\log\mathrm{M}/\mathrm{M}_{\odot} \geq 10.25$, there is a significant positive, monotonic correlation between \deltalblt{} and environment across all six bins. In other words, at $z\geq0.5$ and at fixed stellar mass, galaxies become progressively more bulge-dominated as one moves from the most underdense regions on large scales to the richest cluster environments. For the lower-mass bin at these redshifts, visual inspection of Figure~\ref{fig:deltla_lblt} suggests a similar trend. Under the conservative $5\sigma$ criterion this correlation falls short of confirmation; however, applying BH-FDR does confirm a positive monotonic correlation at $0.5\leq z<0.6$. Table \ref{tab:wilcoxon}, however, shows that when isolating galaxies in cluster environments, the signal is statistically significant in both stellar-mass bins at the $5\sigma$ level: Cluster members are systematically more bulge-dominated than their field counterparts, with significance exceeding the $5\sigma$ level.

%\begin{figure*}[htbp]
%    \centering
%    \includegraphics[width=1\linewidth]{delta_lblt_sfq.png}
%    \caption{NOTE: 3X ERROR BARS}
%    \label{fig:deltla_lblt_sfq}
%\end{figure*}

Unlike the $z<0.5$ case, where morphological changes at fixed stellar mass are only apparent in clusters, the results at $z\geq0.5$ point to a broader role for the environment. For massive galaxies, the positive correlation across all six bins indicates that structural transformations proceed more continuously, with bulge-dominated fractions increasing steadily from voids to rich clusters. For lower-mass galaxies, no such monotonic trend can be confirmed; however, these systems still show clear differences between cluster and field populations at a fixed stellar mass, demonstrating that clusters remain effective sites of transformation across the mass spectrum. Taken together, these results suggest that by $z\geq0.5$, the mechanisms responsible for bulge growth are not confined to the extreme conditions of rich clusters but are already active across a wider range of environments. Processes such as enhanced merger rates, pre-processing in groups or filaments, or interaction-driven bulge growth may therefore operate efficiently in overdense regions beyond the cluster core. The stronger and more widespread environmental dependence at these redshifts is consistent with a picture in which cluster-driven transformations are still ongoing, while overdense environments at large scales also contribute significantly to the buildup of bulge-dominated systems.

We additionally investigated whether the suppressed cluster signal at $z=0.3-0.4$ is compounded by a selection effect. We cross-matched the CAMIRA cluster centers with our galaxy catalog to identify the closest galaxy to each cluster center within $50$\,kpc in the same redshift slice. We find a BCG candidate (median $\log M/M_\odot \sim 11$) in $\sim70\%$ of clusters overall. However, the match rate drops to $63\%$ at $z=0.3-0.4$ and rises to $\sim75\%$ at higher redshifts, indicating that the quality flags (see \S\,\ref{subsec:sample}) might be preferentially removing the brightest cluster members in the lowest redshift bin. Since BCGs are among the most massive and most bulge-dominated galaxies in their clusters, their disproportionate absence at $z=0.3-0.4$ selectively suppresses \deltalblt{} in precisely the high-mass cluster bin where structural saturation already predicts a weakened signal. More broadly, the $63-75\%$ match rates across all redshifts imply that our reported cluster \deltalblt{} values are conservative lower bounds throughout: BCG inclusion would increase the measured offset at every epoch, with the underestimation likely most severe at $z=0.3-0.4$.

A complementary fraction-based presentation of the above results is provided in Appendix~\ref{appendix:morph_fractions}. The trends are fully consistent with the $\Delta L_B/L_T$ analysis above, and are presented separately to enable direct comparison with prior studies that typically report morphological fractions.

\begin{figure}[htbp]
    \centering
    \includegraphics[width=\linewidth]{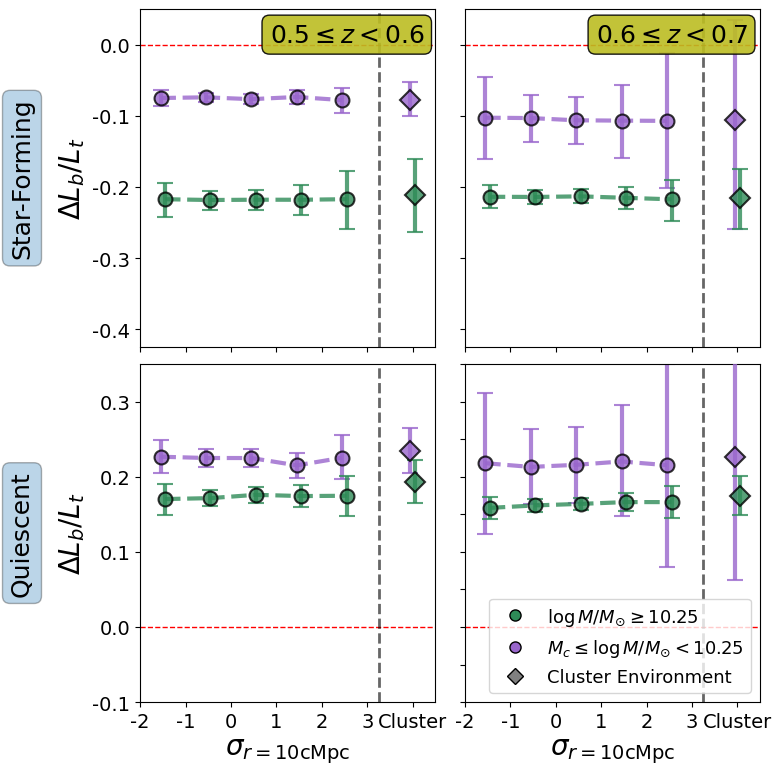}
    \caption{$\Delta L_B/L_T = L_B/L_T - \overline{L_B/L_T}(M)$ is plotted against environmental density separately for star-forming and quiescent sub-populations at $z\geq0.5$. Similar to Figure \ref{fig:deltla_lblt}, the galaxies are also separated into two mass bins with different colors, and the error bars denote the $5\sigma$ uncertainty in the estimation of the median \deltalblt{}.}
    \label{fig:deltla_lblt_sfq}
\end{figure}

\subsubsection{Environmental Trends Within Star-Forming and Quiescent Sub-Populations}

To further probe the $z\geq0.5$ regime, where we see a monotonic increase of median \deltalblt{} across all six density bins, we divide the sample into star-forming and quiescent sub-populations. Figure \ref{fig:deltla_lblt_sfq} shows the variation of \deltalblt{} with environment within each sub-population. The division into star-forming and quiescent sub-populations follows the methodology described in \citet{Ghosh2024}, where galaxies are separated based on their rest-frame SDSS \uband{}-\rb{} versus \rb{}-\zb{} colors, calculated using SED fitting. We refer the interested reader to Appendix B and Figure 8 of \citet{Ghosh2024}. Note that for our sample, we cannot use \textit{UVJ} selection because SED fitting using HSC \grizy{} photometry does not allow robust estimation of rest-frame \textit{J} magnitudes. The SDSS rest-frame magnitudes used for this separation are taken from the same Mizuki catalog employed for redshift and stellar mass estimation. 

As expected, Figure \ref{fig:deltla_lblt_sfq} shows that star-forming galaxies are systematically more disk-dominated, while quiescent galaxies are more bulge-dominated at fixed stellar mass. Within the quiescent sub-population, the higher-mass bin shows a lower $\Delta L_B/L_T$ than the lower-mass bin. Note that this is a natural consequence of the mass-detrending: at high stellar masses, the full-population median $L_B/L_T$ is already elevated, so even bulge-dominated quiescent galaxies deviate less from the baseline than low-mass quiescent galaxies, where the overall population is predominantly disk-dominated.

Remarkably, however, the environmental trends within each sub-population are essentially flat, indicating that $L_B/L_T$ does not vary significantly with large-scale density once stellar mass and star-formation state are controlled for. These flat trends within each sub-population suggest that the environmental processes responsible for the signal seen in Figure \ref{fig:deltla_lblt} are not modifying the bulge-to-disk ratio individually within star-forming or quiescent sub-populations. Rather, the overall correlations in the combined population must therefore arise from associated shifts in the relative abundance of star-forming versus quiescent systems with environment. In other words, the processes driving the aggregate trend appear to act by transforming galaxies from star-forming, disk-dominated systems into quiescent, bulge-dominated ones. This interpretation is consistent with the mechanisms discussed earlier---such as mergers, high-speed tidal encounters, or rapid quenching triggered in dense environments---which are capable of altering both the star-formation state and the morphology of galaxies simultaneously.

\begin{figure*}[htbp]
    \centering
    \includegraphics[width=1\linewidth]{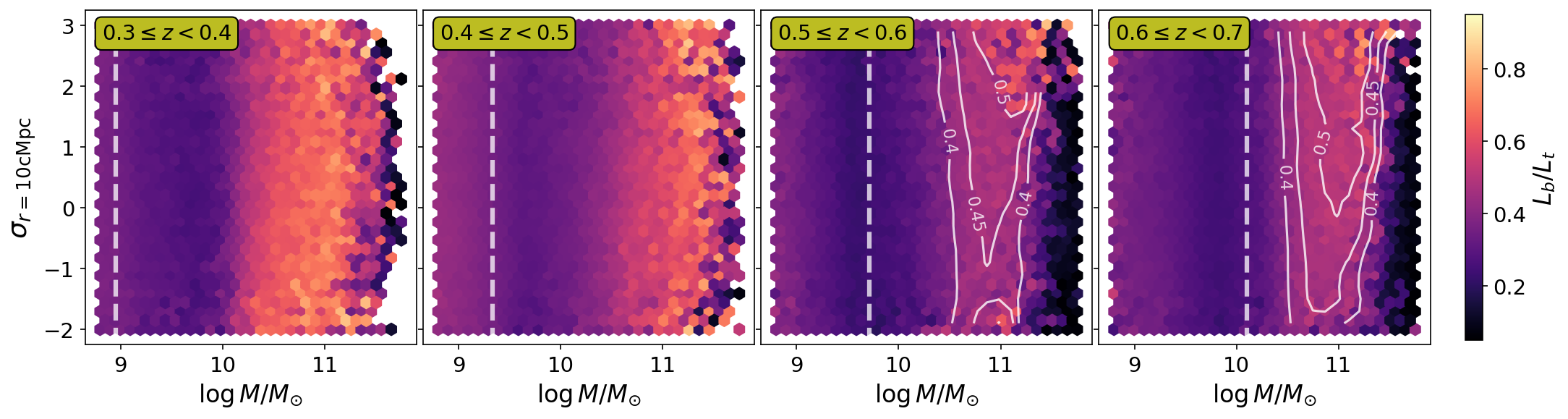}
    \caption{Median bulge-to-total light ratio, across the $\sigma_{r=10\mathrm{cMpc}}$ - stellar mass plane in the four redshift bins. Each hexagonal cell is colored by the median $L_B/L_T$ of galaxies within it (scale shown in color bar). The dashed vertical lines mark the stellar mass completeness limits for each redshift slice. Vertical color gradients (see contours of constant median $L_B/L_T$) at fixed stellar mass indicate a dependence of morphology on large-scale environment, independent of mass. Such gradients are largely absent at $z<0.5$, where a horizontal color gradient dominates, and stellar mass is the primary driver of $L_B/L_T$. At $z\geq0.5$, vertical gradients emerge at high stellar mass, consistent with the monotonic correlations identified in Table \ref{tab:separman}.}
    \label{fig:density_mass_lblt}
\end{figure*}

Figure \ref{fig:deltla_lblt_sfq} also shows that for the cluster points, the positive \deltalblt{} offset seen in Figure \ref{fig:deltla_lblt} is entirely driven by quiescent galaxies. Star-forming galaxies remain offset toward negative \deltalblt{}, indicating no systematic bulge enhancement at fixed stellar mass even in the densest environments. This dichotomy also reflects the coupled transformations between evolutionary states discussed above. Environment increases the fraction of galaxies following the rapid-transformation pathway (green dashed arrow in Figure \ref{fig:gal_evo_schematic}) relative to the secular growth pathway (blue arrow), with morphological transformation and quenching occurring simultaneously rather than sequentially. The increased prominence of the transitional sequence seen at higher redshifts in Figure \ref{fig:lblt_m} suggests that such rapid transformations were more common at earlier epochs. This elevated transformation rate at earlier epochs is consistent with and helps explain the stronger correlation of \deltalblt{} with environment at higher redshifts seen in Figure \ref{fig:deltla_lblt}. 

\subsubsection{Visualizing Morphology on the Environment-Mass Plane}

Figure \ref{fig:density_mass_lblt} offers a complementary qualitative view of the relationship between morphology and large-scale environment by coloring the $\sigma_{r=10\mathrm{cMpc}}$ - stellar mass plane based on the median values of $L_B/L_T$. In this representation, any dependence of morphology on environment at fixed stellar mass would appear as vertical color gradients, with $L_B/L_T$ increasing systematically as one moves upwards in $\sigma_{r=10\mathrm{cMpc}}$. 

The panels demonstrate that such vertical structures are essentially absent at $z<0.5$, regardless of stellar mass, consistent with the lack of monotonic correlations reported in Table \ref{tab:separman}. Instead, the dominant feature is a horizontal gradient, illustrating that stellar mass is the primary driver of $L_B/L_T$ at these redshifts, and the large-scale environment of the galaxy does not play a role. Note that this visualization does not isolate cluster environments separately. As shown earlier, those galaxies do exhibit significant morphological differences at a fixed stellar mass, even though the broader large-scale density field shows little systematic effect.

By contrast, at $z\geq0.5$, the high-mass regime shows a different behavior. In both the high-redshift bins, although the horizontal stellar-mass gradient is still prominent, additional vertical gradients emerge at high stellar masses: median $L_B/L_T$ values increase systematically with $\sigma_{r=10\mathrm{cMpc}}$. To aid in interpretation, the rightmost panels overlay contours of the constant median $L_B/L_T$ on the same plane -- these make the vertical color gradients easier to identify. This trend is fully consistent with the significant positive correlations reported for the higher mass bin in Table \ref{tab:separman}. Figure \ref{fig:density_mass_lblt} provides an alternative visualization of the quantitative results, highlighting where in mass-redshift-environment space the influence of large-scale density on morphology is most apparent.

We note that the anomalously low $L_B/L_T$ colors at the extreme high-mass end ($\log M/M_\odot \gtrsim 11.5$) in Figure \ref{fig:density_mass_lblt} are a sampling artifact: there are very few galaxies at those masses, resulting in $\sim1-2$ galaxies per hexbin. Therefore, the median $L_B/L_T$ values here are unreliable and should not be interpreted as a physical trend.

\section{Discussion} \label{sec:conclusion} 
In this section, we discuss the physical context and implications of the results presented in \S\ref{sec:results} and compare our results against prior work. We emphasize that our data directly constrain only stellar morphology; the physical mechanisms discussed below are inferred by analogy with prior observational and theoretical work rather than being directly tested by our analysis. In future work, we aim to directly compare our results with hydrodynamic simulations to more rigorously test the proposed mechanisms. Here, we instead situate our findings within the observational and theoretical constraints available in the literature.
 
\begin{figure*}[htbp]
    \centering
    \includegraphics[width=0.6\linewidth]{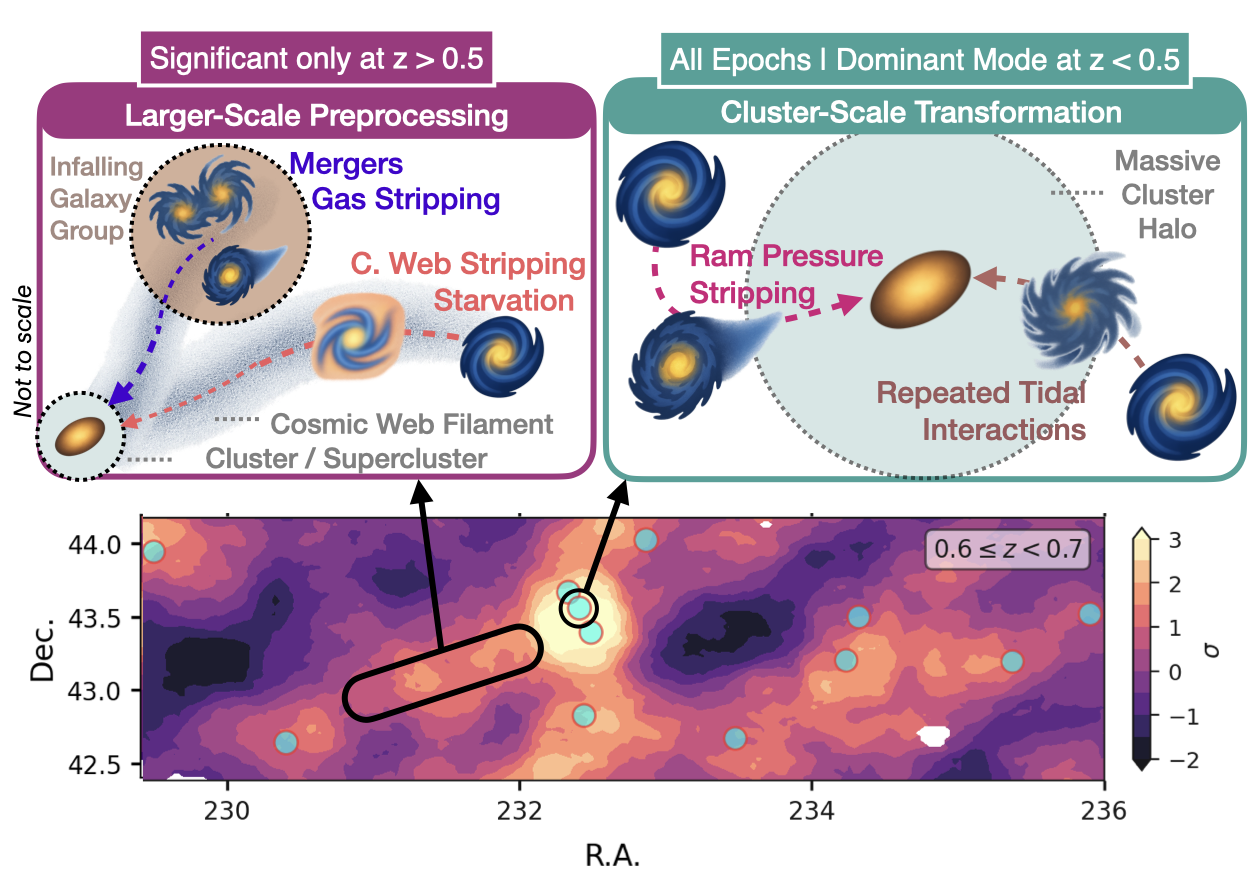}
    \caption{Schematic illustration of dominant environmental mechanisms driving structural transformation at different cosmic epochs, shown on an HSC overdensity map from Figure \ref{fig:density_map}. (\textit{Right Panel}): At $z < 0.5$, cluster-specific processes such as ram-pressure stripping and tidal interactions dominate structural transformation within $\sim2$ cMpc scales. (\textit{Left Panel}): At $z \geq 0.5$, transformation operates across broader scales through mergers, gas stripping, and starvation in groups and filaments, in addition to cluster scale transformation.}
    \label{fig:gal_evo}
\end{figure*}

\subsection{Implications for Galaxy Evolution}
 The results presented in \S\ref{sec:results} provide interesting insights into the physical mechanisms driving galaxy structural evolution at different stellar mass and environmental scales. The distinct behavior between redshift regimes also points to fundamental differences in the dominant environmental processes responsible for the structural transformation of galaxies at these different epochs. Our findings challenge simplified galaxy formation models that treat environment only as a binary (cluster/field) or assume universal environmental effects across cosmic time and stellar mass regimes. Figure \ref{fig:gal_evo} provides a simplified schematic summary of how the dominant environmental mechanisms vary with cosmic epoch and spatial scale, as suggested by our analysis and discussed in the following subsections. 

\subsubsection{Variation of Dominant Environmental Mechanisms with Redshift} \label{subsec:var_with_z}
The cluster-only signal at $z<0.5$ suggests that extreme environmental processes operating at cluster scales --- such as ram pressure stripping and repeated tidal interactions (also called ``harassment") --- dominate structural transformation at these epochs. For example, ram pressure stripping becomes effective only when the ram pressure from the intracluster medium exceeds the gravitational restoring force that holds the galaxy's gas in place -- a condition typically satisfied only in cluster cores \citep[e.g.,][]{Gunn1972OnEvolution, Boselli2022RamEnvironments}. Similarly, repeated tidal interactions are most effective at transforming intermediate mass galaxies, with the efficiency rising on eccentric orbits and at smaller cluster-centric radii \citep[e.g.,][]{Moore1996GalaxyGalaxies,Bialas2015OnHarassment}. Therefore, these mechanisms require the high relative velocities and dense intracluster medium found only in massive clusters and provide a physically motivated explanation for why we detect structural enhancement preferentially in the richest environments at late epochs. A notable exception is seen for the most massive galaxies in our lowest redshift slice, where clusters no longer produce a discernible offset at a fixed stellar mass. We interpret this as a regime of structural saturation: by $z \sim 0.3$, 
earlier-epoch transformations have rendered the most massive galaxies predominantly bulge-dominated across all environments, leaving little residual morphological contrast between cluster and field members at fixed stellar mass. As discussed in 
\S\,\ref{subsec:lblt_v_env}, the preferential exclusion of BCGs from our sample at $z=0.3$--$0.4$ may additionally contribute to the suppressed cluster offset in this bin, since BCGs are often the most bulge-dominated members of their clusters. This cluster-dominated regime is illustrated in the right panel of Figure \ref{fig:gal_evo}.

By contrast, the broader environmental correlation we observe at $z\geq0.5$ indicates additional mechanisms operating across a wider range of environmental densities. The additional non-cluster effects associated with this correlation could possibly include enhanced merger rates in moderately overdense regions \citep[e.g.,][]{Fakhouri2010TheSimulations, Lin2010WHERESURVEY, Pearson2024EffectsFraction} as well as preprocessing in galaxy groups and filaments prior to cluster infall and potential cosmic-web stripping that can remove gas \citep[e.g.,][see \S \ref{subsec:preproc}]{McGee2009TheClusters, Dressler2013THEASSEMBLY}. These broader environmental effects, operating across filaments and groups as illustrated in the left panel of Figure \ref{fig:gal_evo}, augment the cluster-specific processes that continue to operate at these epochs.

Our results also reflect the maturation of environmental mechanisms as cosmic structures developed and intracluster medium densities increased --- it points to the joint evolution of halo gas content, satellite orbits, and galaxy gas reservoirs. At earlier epochs, galaxies were on average more gas-rich, merger rates were higher, and group-scale preprocessing was more common \citep[e.g.,][]{Conselice2009The1.2, McGee2009TheClusters, Lotz2011THE1.5, Kawinwanichakij2017Effectsup/sup, Tacconi2020TheTime}, making galaxies more susceptible to structural transformations at a wide variety of environmental scales. By $z<0.5$, the expected depletion of cold gas in galaxies and the buildup of dense intracluster media are consistent with ram pressure stripping and tidal interactions becoming more effective and the dominant environmental mechanism for structural transformation.

Finally, the increased prominence of transitional systems in Figure \ref{fig:lblt_m} at higher redshifts indicates that structural transformation events were more frequent at earlier epochs. This is consistent with the elevated merger rates at earlier epochs discussed above. The overall implication is that even moderately dense environments at earlier epochs could act as accelerators of structural transformation, whereas in the more mature cosmic web at lower redshifts, only the most extreme cluster environments retain this efficiency.

\subsubsection{Preprocessing and Multi-Scale Effects} \label{subsec:preproc}
As indicated in \S \ref{subsec:var_with_z}, the detection of morphological differences across the full range of environments at $z \geq 0.5$ provides evidence that transformation is not confined to rich clusters at these epochs, but also occurs in galaxy groups and along cosmic-web filaments before cluster infall. Such preprocessing implies that a significant fraction of galaxies have already been structurally altered prior to their first significant encounter with a massive cluster potential. Consistent with this picture, simulations suggest that group preprocessing affects a substantial fraction of cluster galaxies before their arrival in massive clusters \citep[e.g.,][]{Fujita2004Pre-ProcessingCluster,McGee2009TheClusters, Donnari2020QuenchedPre-processing}.

Moreover, groups at these epochs were more gas-rich and dynamically active, making them efficient sites for mergers, tidal encounters, and gas stripping \citep[e.g.,][]{Zabludoff1998HierarchicalGalaxies, McGee2009TheClusters}. In addition, the filamentary structures feeding clusters provide further opportunities for preprocessing through multiple pathways: enhanced merger rates and tidal interactions can trigger morphological transformations and starbursts, while slower processes such as starvation or cosmic-web stripping gradually deplete cold gas reservoirs before cluster infall. Observations of cosmic filaments have revealed elevated star formation rates, increased fractions of mergers and starbursts, as well as morphological disturbances in filament galaxies compared to the field \citep[e.g.,][]{Edwards2010iSPITZER/iDATA,  AragonCalvo2019GalaxyDetachment, Sarron2019Pre-processingCFHTLS, Edwards2021EfficientSITELLE, Barsanti2022TheBulge}. These pathways are consistent with the significant morphological differences we detect even in moderately overdense regions at higher redshift.

We should also note that the physical mechanisms operating during preprocessing differ qualitatively from those in massive clusters. These processes (e.g., gradual removal of warm gas halos that serve as reservoirs for future star formation) are typically slower than the rapid ram pressure stripping characteristic of clusters but more efficient than field evolution. This is consistent with the higher value of \deltalblt{} in clusters (compared to the $\sigma_{r=10\mathrm{cMpc}}$ points) in Figure \ref{fig:deltla_lblt}. 

This multi-scale environmental influence suggests that galaxy transformation is not a simple function of instantaneous local density but instead reflects a complex interplay between accretion history, halo mass growth, and cumulative environmental exposure. Our results suggest that theoretical studies and semi-analytic models must therefore track not only the instantaneous environment of galaxies but also their integrated environmental history: the time spent as satellites in groups, the frequency of passages through filaments, and the trajectories by which galaxies are accreted onto clusters.

We note an important limitation of our large-scale density tracer in this context. The $r = 10\,\mathrm{cMpc}$ aperture used to compute density in this work is considerably larger than typical group virial radii, and therefore cannot isolate individual preprocessing environments. However, because galaxy groups cluster along filaments and in moderately overdense nodes, galaxies preprocessed in groups will on average reside in regions of elevated $\sigma_{r=10\,\mathrm{cMpc}}$. Our density field thus captures the statistical aggregate signature of preprocessing rather than resolving individual group membership, and cannot cleanly disentangle preprocessing contributions from other large-scale effects such as cosmic-web stripping or enhanced merger rates in filaments.

\subsubsection{Coupled Environmental Quenching and Morphological Transformation}

Figure \ref{fig:deltla_lblt_sfq} shows that, at $z\geq0.5$, the environmental signal does not manifest as structural variation within fixed star-formation-state populations but rather as shifts in the relative fractions of star-forming and quiescent galaxies, consistent with environment driving transitions between distinct evolutionary states. Star-forming galaxies show systematically negative \deltalblt{} values (more disk-dominated) regardless of environment, while quiescent galaxies show positive values (more bulge-dominated), with the environmental trend emerging from shifts in the relative fractions of these populations. The lack of structural variation within each population supports a model where environmental quenching and morphological transformation are coupled processes, with gas removal mechanisms operating alongside stellar mass assembly processes \citep[e.g., compare Figures \ref{fig:lblt_m} and \ref{fig:ssfr_lblt_mass}; also see][]{schawinski_14, Peng2015StrangulationGalaxies}.

The environmental effect operates by increasing the fraction of galaxies following rapid transformation pathways (Figure \ref{fig:gal_evo_schematic}, green dashed arrow) relative to secular evolution (blue arrow). Additionally, in dense environments, galaxies are more likely to experience processes - such as major mergers, violent disk instabilities, or rapid gas removal combined with tidal heating - that simultaneously quench star formation and restructure the stellar distribution \citep[][]{Dekel2009FORMATIONSPHEROIDS, Hopkins2009TheDemographics, Rodriguez-Gomez2017TheMorphology}.

The timescales for these coupled transformations likely vary with the specific physical process: Violent events like major mergers can simultaneously quench star formation and restructure galaxies within $\sim1$ Gyr \citep[e.g.,][]{Wild2009Post-starburstCuriosity, schawinski_14}, while more gradual processes may operate over several Gyr but still maintain the coupling between structural and star formation changes \citep[e.g.,][]{Wetzel2013GalaxyUniverse, Peng2015StrangulationGalaxies}. This framework is consistent with the distinct evolutionary tracks in the $L_B/L_T$-stellar mass plane we observe.

\subsection{Comparison With Previous Studies}\label{sec:lit_comp}
Our results help reconcile seemingly conflicting prior results on this topic (e.g., see Table \ref{tab:lit_survey}) by demonstrating that the morphology-environment correlation at a fixed stellar mass is neither universally present nor universally absent. Once the dominant mass trend is removed and multi-scale environments are considered with adequate statistics and uncertainty propagation, residual environment-structure correlations emerge, but their detectability and strength depend on where one sits in the (z, $M_*$, environmental-scale) space. This naturally explains why earlier works, each sampling different slices of that space with much smaller samples, reached divergent conclusions sometimes.

Among the studies described in Table \ref{tab:lit_survey}, \citet{EuclidCollaboration2025Euclidz=1} provides the most relevant comparison, as it investigates a similar redshift regime with the largest sample size. They report that at a fixed stellar mass, the fraction of early-type galaxies increases with environmental density, primarily out to $z\sim0.75$ and for $\log\mathrm{M}/\mathrm{M}_\odot \leq 10.8$. We should note that there are a few important differences between the two studies. A key methodological difference is the definition of ``morphology". To study the rest-frame $B$ band at these redshift ranges, \citet{EuclidCollaboration2025Euclidz=1} would have to use the $i$ and $z$ bands, which Euclid does not observe. Therefore, the authors classify early-type galaxies using a demarcation in the \sersic{} index - color plane, rather than a purely structural measurement. By construction, this classifier is biased toward red systems and is not exclusively based on visual/structural features. In contrast, we adopt a strictly structural definition of morphology at an equivalent rest-frame wavelength of $450\,\mathrm{nm}$ across all redshifts.

Because the Euclid early-type classification blends color with structure, its morphology-density relation partially reflects the color-density relation by design. This distinction matters most in environments where star formation and structure evolve on different timescales. For example, in dense environments, many disks might be quenched by gas removal (e.g., ram pressure stripping, strangulation) before they significantly grow a bulge component. These galaxies become red but remain disk-dominated structurally. However, a color-inclusive classification scheme will count such disks in their ``early-type" fraction, potentially inflating environmental trends. Given that gas-removal processes are the dominant environmental structural-transformation mechanism at lower redshifts, we expect color-inclusive classifiers to report a stronger apparent environmental trend at lower redshifts. 

The above difference might explain why the two studies find environmental signatures with similar overall patterns but different detailed amplitudes and mass/redshift boundaries. We additionally note that our work uses a sample that is an order of magnitude larger, $\sim0.5$ dex deeper in mass-completeness, and has access to well-calibrated uncertainties. These factors enhance our ability to detect subtle correlations and properly quantify their significance. However, importantly, both studies converge on the same fundamental conclusion: Environmental effects on galaxy structure exist at fixed stellar mass but are secondary to the dominant role of stellar mass itself. The broad agreement between independent datasets, analysis techniques, and morphological indicators strengthens confidence that the reported overall correlation represents a genuine, if subtle, physical process rather than a methodological artifact. 

Looking ahead, applying \gampen{}'s probabilistic framework to Euclid imaging would combine Euclid's superior spatial resolution with robust uncertainty quantification, potentially enabling detection of subtle environmental effects at the statistical rigor achieved here but extending to higher redshifts.

 \section{Summary and Conclusions} \label{sec:summary}

 In this work, we presented a comprehensive analysis of the variation of galaxy structure with environment beyond the local universe, while disentangling it from the impact of stellar mass. Our sample comprises $\sim3$ million HSC galaxies with $m<23$ ($\equiv \log M/M_{\odot} \geq 8.9$ at the lowest redshifts) spanning $0.3 \leq z < 0.7$. Our structural parameter measurements are anchored to a consistent effective rest-frame wavelength of $\sim450\,\mathrm{nm}$ and are accompanied by well-calibrated posterior distributions from the \gampen{} framework. Our environmental probe combines large-scale projected overdensities measured in $r=10\,\mathrm{cMpc}$ apertures with cluster membership from the CAMIRA catalog (association radius $\sim2\,\mathrm{cMpc}$), thereby probing both the large-scale cosmic web as well as rich clusters. Compared to most previous studies at comparable redshifts, this dataset provides a $1$ - $4$ dex improvement in sample size and at least a $\sim0.5$ dex improvement in mass completeness.

To isolate environmental effects at fixed stellar mass, we defined $\Delta L_B/L_T = L_B/L_T - \overline{L_B/L_T}(M)$, and carried out a Bayesian correlation analysis that propagates full posterior uncertainties through every step. This construction removes the mass dependence of morphology while allowing us to fully incorporate uncertainties in our structural parameter measurements, providing a stringent test of whether galaxy structure is correlated with environment when controlling for stellar mass and thereby addressing a fundamental question that has persisted since \citeauthor{dressler1980}'s pioneering work in \citeyear{dressler1980}.

\subsection{Primary Findings}
Our primary result definitively establishes:
\begin{quote}
Galaxy structure depends on environment even after controlling for stellar mass, but this dependence is secondary to stellar mass; the strength of this correlation varies with a) the environmental scale being probed; b) redshift; and c) stellar mass regime.
\end{quote}

When the large-scale and cluster-scale statistical tests are considered together (Tables \ref{tab:separman} and \ref{tab:wilcoxon}), we can confirm with more than $5\sigma$ confidence the presence of a correlation between $L_B/L_T$ and environment at a fixed stellar mass across almost all redshift and stellar mass bins. The monotonic trends detected by our statistical tests are modest in amplitude and only reach formal statistical significance given the sample size. Thus, structure-environment correlation exists independent of stellar mass but with amplitudes well below the primary structure-stellar mass relationship (compare Figure \ref{fig:lblt_m} with Figure \ref{fig:deltla_lblt}). 

At $z<0.5$, we find no statistically significant monotonic correlation between $L_B/L_T$ and large-scale ($10$ cMpc) environmental density when stellar mass is controlled for. However, galaxies in rich clusters at these redshifts show statistically significant morphological differences, with cluster members being systematically more bulge-dominated ($\Delta L_B/L_T > 0$ at $>5\sigma$ significance) compared to similarly massive field galaxies. This indicates that at these epochs, environmental transformation is primarily driven by cluster-specific processes rather than large-scale overdensity.

This picture changes considerably at $z\geq0.5$. For massive galaxies ($\log\mathrm{M}/\mathrm{M}_\odot\geq10.25$), we detect a positive monotonic correlation between \deltalblt{} and environment across all six environmental bins with $>5\sigma$ confidence. This correlation extends from cosmic voids to rich clusters, suggesting that environmental mechanisms capable of driving structural transformation operate across broader spatial scales at earlier cosmic epochs. For lower-mass galaxies at $z\geq0.5$, the large-scale monotonic trend falls below the conservative $5\sigma$ threshold, though BH-FDR confirms a positive correlation at $0.5\leq z<0.6$. Cluster members across the full lower-mass range remain significantly more bulge-dominated compared to field galaxies at the same stellar mass. This demonstrates that clusters remain effective sites of transformation across almost the entirety of the mass and redshift regime covered in this study. 

%At $z<0.5$, stellar mass dominates morphological trends with minimal environmental dependence, except cluster-specific differences. At $z\geq0.5$, both disk-dominated and bulge-dominated fractions clearly show monotonic trends with environment, though this remains secondary to the strong mass dependence.
Possible physical mechanisms underlying these trends -- including structural saturation in the low-redshift cluster regime, preprocessing in groups and filaments at earlier epochs, and the coupling between environmental quenching and morphological transformation -- are discussed in depth in \S\ref{sec:conclusion}.

\subsection{Conclusions and Future Direction}
This study marks a significant advance in addressing the relative roles of stellar mass and environment in shaping galaxy structure beyond the local universe. Using an unprecedentedly large sample, we demonstrate that environment does influence galaxy structure beyond what stellar mass alone predicts, but this influence is neither universal nor uniform. Instead, it varies systematically with cosmic time, environmental scale, and galaxy mass in ways that reveal the underlying physical processes driving galaxy evolution.

Our results also underscore that understanding galaxy evolution requires careful consideration of the multi-dimensional parameter space of galaxy physical properties as well as their environmental characteristics. Furthermore, the methodological framework presented in this work provides a versatile tool for probing the interplay between galaxy properties and multi-scale environment. This approach will be particularly valuable in light of ongoing and forthcoming large spectroscopic surveys, such as the Dark Energy Spectroscopic Instrument \citep[DESI;][]{DESICollaboration2016} and the Subaru Prime Focus Spectrograph \citep[PFS;][]{Takada2014}. These initiatives will provide the high-resolution data needed to examine structural-environment correlations across extensive cosmic volumes, enabling analyses that transcend redshift slicing. In addition, future imaging and grism data from the Vera C. Rubin Observatory Legacy Survey of Space and Time \citep[Rubin-LSST;][]{Ivezic2019LSST:Products}, Euclid and the Nancy Grace Roman Space Telescope \citep[NGRST;][]{Spergel2013, ngrst} promise to extend these investigations at similar statistical significance to lower masses and higher redshifts, facilitating a deeper understanding of the evolution of these relationships over cosmic time.

\appendix

\section{Selected Previous Studies of Morphology-Environment Correlation} \label{sec:ap:lit_survey}

Table~\ref{tab:lit_survey} summarizes a representative selection of observational studies examining the correlation between galaxy morphology and environment, spanning from early work in the local universe to more recent investigations at intermediate and high redshifts. For each study, we record the survey or instrument used, the redshift range and sample size, the density measure employed, and, crucially, whether the study finds a morphology-environment correlation that persists after controlling for stellar mass (final column). The final row includes the present work for comparison; see \S\ref{sec:lit_comp} for a detailed discussion of how our results relate to prior studies.

\begin{deluxetable*}{P{2.25cm}P{2cm}cccP{1.75cm}P{2.5cm}P{3.25cm}}[htbp]
%\tablenum{2}
\tablecaption{\textit{Selected} Studies of Morphology-Environment Correlation\label{tab:lit_survey}}
\tablecolumns{7}
\tablehead{
\colhead{Reference} & \colhead{Survey/} & \colhead{Redshift} & \colhead{Mass/} & \colhead{Sample} & \colhead{Density} & \colhead{Correlation with} & \colhead{Correlation at}\\
\colhead{} & \colhead{Instrument\tablenotemark{a}} & \colhead{} & \colhead{Magnitude\tablenotemark{b}} & \colhead{Size} & \colhead{Measure}& \colhead{Environment\tablenotemark{c}} & \colhead{Fixed $M_*$\tablenotemark{c}}}
\startdata
    \hline
    \hline
    \multicolumn{8}{c}{Early Studies at $z\leq0.2$}\\
    \hline
    \citeauthor{dressler1980} \citeyear{dressler1980} & Las Campanas & $\leq0.06$ & - & $\sim 6000$ & Cluster/ Field& S0/Elliptical: $\textcolor{teal}{\pmb{\uparrow}}$ Spirals: $\textcolor{red}{\pmb{\downarrow}}$ & Not performed\\
    \hline
    \citeauthor{Kauffmann2004TheGalaxies} \citeyear{Kauffmann2004TheGalaxies} & SDSS & $\leq0.1$ & $\geq2\times10^9$ & $\sim 17,000$ & Density Contrast & & Concentration: $\textcolor{teal}{\pmb{\uparrow}}$ only at $\leq10^{10.5}\mathrm{M}_{\odot}$\\
    \hline
    \citeauthor{vanderWel2008} \citeyear{vanderWel2008} & SDSS & $0.02-0.03$ & $\geq 10^{10}$ & $\sim4500$ & Density Contrast & & \sersic{} Index: weak $\textcolor{teal}{\pmb{\uparrow}}$\\
    \hline
    \citeauthor{Bamford2009} \citeyear{Bamford2009} & SDSS & $0.03-0.085$ & $\geq 10^{9.5}$ & $\sim2\times10^5$ & Density Contrast & & Elliptical: weak $\textcolor{teal}{\pmb{\uparrow}}$\\
    \hline
    \hline
    \multicolumn{8}{c}{Studies at $z>0.2$}\\
    \hline
    \citeauthor{Dressler1997EvolutionGalaxies} \citeyear{Dressler1997EvolutionGalaxies} & HST & $\sim0.5$ & $R\leq23$ & $\sim 1900$ & Cluster/ Field & S0/Elliptical $\textcolor{teal}{\pmb{\uparrow}}$ Spirals: $\textcolor{red}{\pmb{\downarrow}}$ & Not performed\\
    \hline
    \citeauthor{Postman2005TheClusters} \citeyear{Postman2005TheClusters} & HST & $0.8-1.2$ & $I\leq23.5$ & $\sim5000$ & Cluster/ Field & S0/Elliptical: $\textcolor{teal}{\pmb{\uparrow}}$ Spirals: $\textcolor{red}{\pmb{\downarrow}}$ & Not Performed\\
    \hline
    \citeauthor{vanderWel2007TheGalaxies} \citeyear{vanderWel2007TheGalaxies} & GOODS-S & $0.6-1.0$ & $\geq 4\times10^{10}$ & $\sim150$ & Density Contrast & Early Type: $\textcolor{teal}{\pmb{\uparrow}}$ & Not Performed\\
    \hline
    \citeauthor{Huertas-Company2009TheClusters} \citeyear{Huertas-Company2009TheClusters} & MegaCam & $0.4-0.6$ & $\geq10^{9.5}$ & $\sim4000$ & Cluster/ Field &  & Early-Type: $\textcolor{teal}{\pmb{\uparrow}}$\newline Late-Type: $\textcolor{red}{\pmb{\downarrow}}$ \newline only at $\geq10^{10.3}\,\mathrm{M}_\odot$\\
    \hline
    \citeauthor{Tasca2009} \citeyear{Tasca2009} & zCOSMOS & $0.2-1.0$ & $\geq 10^{10}$ & $\sim10,000$ & Density Contrast &  & Early Type: weak $\textcolor{teal}{\pmb{\uparrow}}$ only at $\leq10^{10.6}\mathrm{M}_{\odot}$\\
    \hline
    \citeauthor{Kovac2010THE1} \citeyear{Kovac2010THE1} & zCOSMOS & $0.1-1.0$ & $\geq10^{9.9}$ & $\sim8,500$ & Density Contrast &  & Early Type: weak $\textcolor{teal}{\pmb{\uparrow}}$ only at $z\leq0.4$\\
    \hline
    \citeauthor{Kawinwanichakij2017Effectsup/sup} \citeyear{Kawinwanichakij2017Effectsup/sup} & ZFOURGE & $0.5-2.0$ & $\geq10^{9}$ & $\sim5000$ & Density Contrast &  & \sersic{} Index: $\textcolor{teal}{\pmb{\uparrow}}$ only at $0.5<z<1.0$ and $\gtrsim10^{10.25}\mathrm{M}_\odot$\\
    \hline
    \citeauthor{Sazonova2020TheClusters} \citeyear{Sazonova2020TheClusters} & HST & $1.2-1.8$ & $H \leq 24$ & $\sim 300$ & Cluster/ Field &  & \sersic{} Index: $\textcolor{teal}{\pmb{\uparrow}}$ for $50\%$ of clusters\\
    \hline
    \citeauthor{Chan2021The1.4} \citeyear{Chan2021The1.4} & HST & $1.0-1.3$ & $\geq10^{9.5}$ & $\sim800$ & Cluster/ Field & & Axis Ratio: $\textcolor{red}{\pmb{\downarrow}}$ only for quiescent subsample at $10^{10.1-10.5}\mathrm{M}_\odot$ \\
    \hline
    \citeauthor{Gu2021TheCANDELS} \citeyear{Gu2021TheCANDELS} & CANDELS & $0.5-2.5$ & $10^{9.2}$ & $\sim14,000$& Density Contrast & & \sersic{} Index: $\textcolor{teal}{\pmb{\uparrow}}$ only at $z<1.0$ and $\geq10^{10}\mathrm{M}_\odot$\\
    \hline
    \citeauthor{Shimakawa2021} \citeyear{Shimakawa2021} & HSC & $0.3-0.6$ & $\geq5\times10^{10}$ & $\sim2.7\times10^5$& Density Contrast & Spirals: $\textcolor{red}{\pmb{\downarrow}}$ & Not Performed\\
    \hline
    \citeauthor{Mei2023Morphology-density2.8} \citeyear{Mei2023Morphology-density2.8} & CLARA & $1.4-2.8$ & $\geq10^{9.6}$  & $\sim250$ & Cluster/ Field & Early Type:$\textcolor{teal}{\pmb{\uparrow}}$ & Not Performed\\
    \hline
    %\citeauthor{Strazzullo2023GalaxyClusters} \citeyear{Strazzullo2023GalaxyClusters} & SPT/HST & $1.4-1.7$ & $\geq10^{10.85}$ & $\sim100$ & Cluster/ Field & \sersic{} Index: $\textcolor{teal}{\pmb{\uparrow}}$ overall; \ding{53} for quiescent sample & Not Performed\\
    %\hline
    \citeauthor{EuclidCollaboration2025Euclidz=1} \citeyear{EuclidCollaboration2025Euclidz=1} & Euclid Q1 & $0.25-1$ & $\geq10^{9.5}$ & $\sim5\times10^5$ & Density Contrast & & Early Type\tablenotemark{*}: $\textcolor{teal}{\pmb{\uparrow}}$ only at $z\leq0.75$ and $\leq10^{10.8}\mathrm{M}_{\odot}$\\
    \hline
    \textit{This work} & HSC & $0.3-0.7$ & $\geq10^{8.9}$ & $\sim3\times10^6$ & Density Contrast $+$ Cluster/ Field & & $\Delta L_B/L_T$:\hspace{1cm} Cluster: $\textcolor{teal}{\pmb{\uparrow}}$ at all z, M Large-Scale: $\textcolor{teal}{\pmb{\uparrow}}$ only at $z\geq0.5; \geq10^{10.25}\,\mathrm{M}_\odot$\\
\enddata
\tablenotetext{a}{For studies that used multiple data-sources or instruments; the primary source/instrument is mentioned in the table}
\tablenotetext{b}{When mentioned in powers of $10$, the limit is referring to the stellar-mass completeness/threshold in units of $\mathrm{M}_\odot$. When mass completeness information is not available, we provide limiting AB magnitudes with the relevant band mentioned.}
\tablenotetext{c}{$\textcolor{teal}{\pmb{\uparrow}}$ (positive correlation) refers to the fraction of galaxies of the specified class or the specified parameter increasing in denser environments, while $\textcolor{red}{\pmb{\downarrow}}$ (negative correlation) indicates a corresponding decrease.}
\tablenotetext{*}{Early Type classification uses color information; not exclusively structural}
%\tablecomments{}
\end{deluxetable*}

\section{Stellar Mass Distributions Across Environment Bins} \label{appendix:cdfs}

\begin{figure*}[h!]
    \centering
    \includegraphics[width=\linewidth]{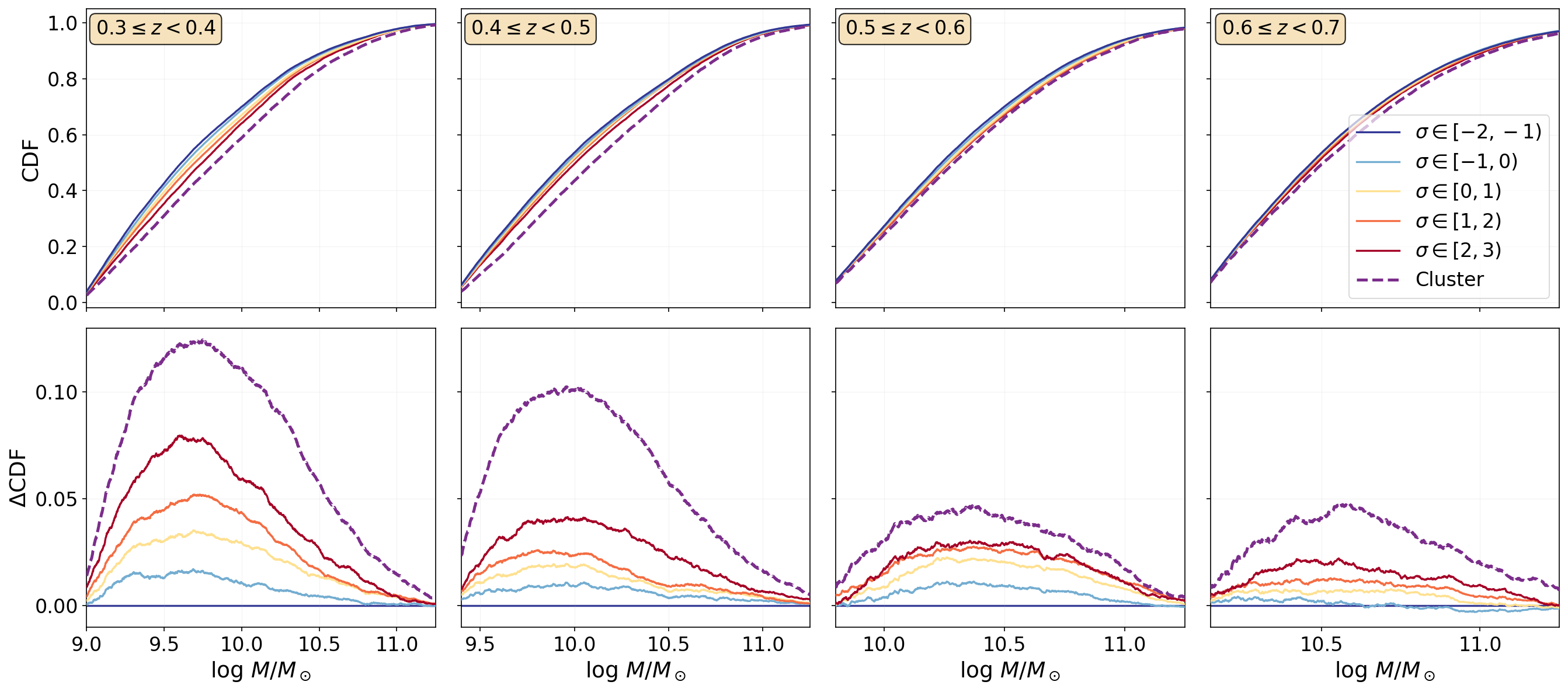} 
    \caption{(\textit{Top:}) Cumulative distributions of stellar mass for galaxies in each of the six environment bins across the four redshift slices. (\textit{Bottom:}) The difference in CDF relative to the most underdense bin ($\Delta\mathrm{CDF} = \mathrm{CDF}(\sigma\in[-2,-1)) - \mathrm{CDF}(\mathrm{bin}\,i)$) is shown for all four redshift slices. Positive values indicate that a given bin contains proportionally more massive galaxies than the underdense reference.}                                                    
  \label{fig:mass_cdf_env}
\end{figure*}

Figure~\ref{fig:mass_cdf_env} shows the cumulative stellar mass distributions for each of the six environment bins across all four redshift slices. As can be seen from the lower panel, the $\Delta$CDF increases monotonically from underdense to cluster environments across all redshift slices, confirming that stellar mass distributions differ systematically with environment and motivating the $\Delta L_B/L_T$ mass-detrending approach used throughout this work. A Kolmogorov--Smirnov test confirms that these differences are statistically significant in all redshift slices (KS $p < 10^{-10}$ in all but two large-scale bins at $z \geq 0.6$). The magnitude of the effect, measured by the peak $\Delta$CDF between the underdense and cluster bins, decreases systematically with redshift: from $\sim$0.12 at $0.3\leq z<0.4$ to $\sim$0.05 at $0.6\leq z<0.7$.  We note that this apparent decrease is at least partly a selection effect, as the rising mass completeness threshold at higher redshifts compresses the accessible stellar mass range and reduces the dynamic range over which environmental differences can manifest.

\section{Mass-Split Violin Distributions} \label{appendix:mass_split_violins}

\begin{figure*}[htb]
    \centering
    \includegraphics[width=1.0\linewidth]{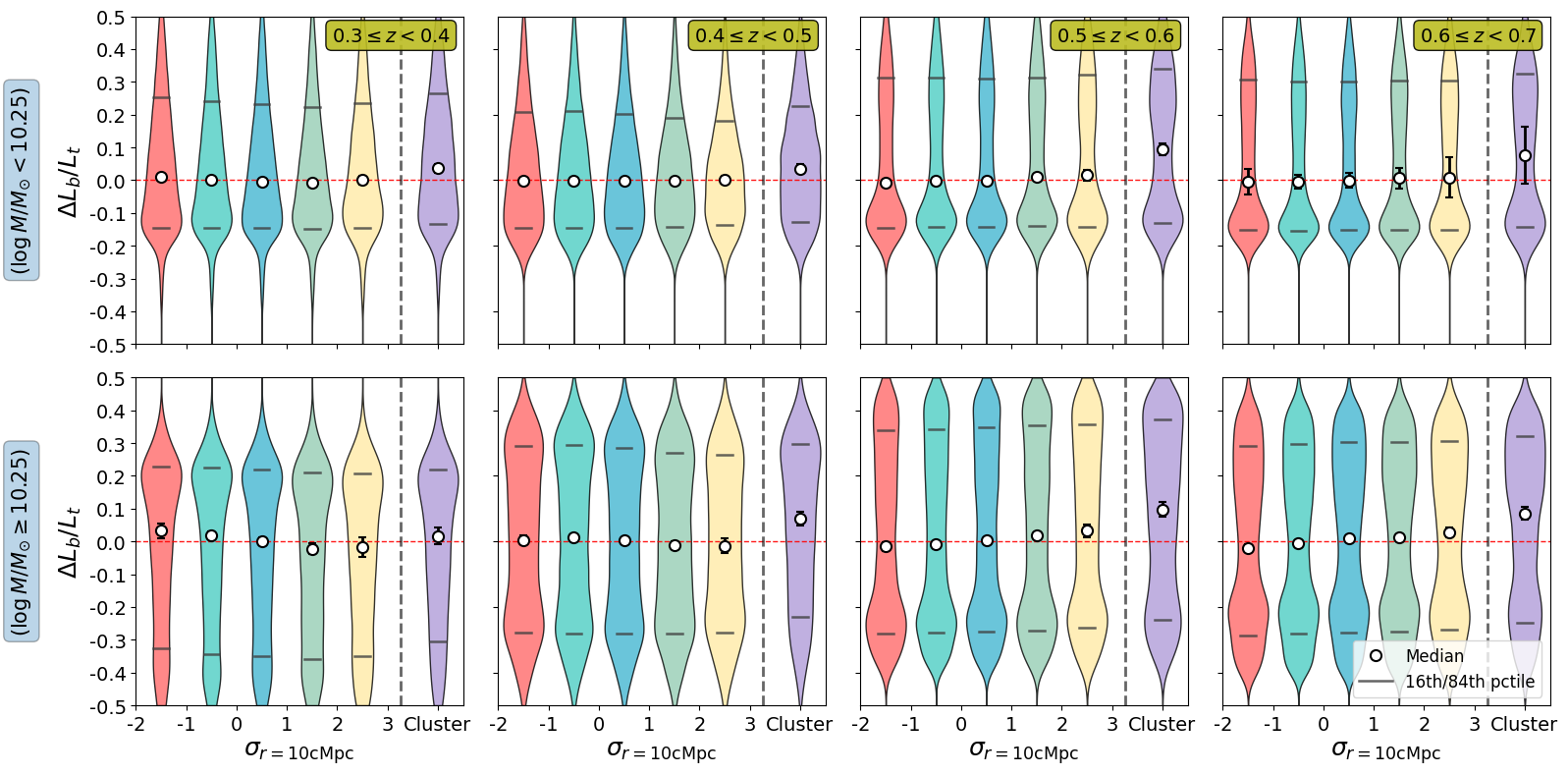}
    \caption{Similar to Figure \ref{fig:violin_all} but with the total sample split into two stellar mass bins. The upper row shows all galaxies with $\log \mathrm{M}/\mathrm{M}_{\odot} < 10.25$; while the lower row shows more massive galaxies with $\log \mathrm{M}/\mathrm{M}_{\odot} \geq 10.25$. The chosen threshold of $10^{10.25}\mathrm{M}_{\odot}$ corresponds to the $\sim75^{th}$ percentile of the overall stellar mass distribution. The horizontal bars demarcate the 16$^{\mathrm{th}}$--84$^{\mathrm{th}}$ percentile range of each distribution. The white points in each violin mark the median $\Delta L_B/L_T$, with error bars showing the $5\sigma$ uncertainty on the median estimate. Splitting the sample by stellar mass shows that large-scale environmental trends emerge only for massive ($\log \mathrm{M}/\mathrm{M}_\odot \geq 10.25$) galaxies at $z\geq0.5$, while cluster-driven bulge enhancement is present in both mass bins.}
    \label{fig:violin_2mass_bins}
\end{figure*}

Figure \ref{fig:violin_2mass_bins} presents the \deltalblt{} violin distributions from Figure \ref{fig:violin_all}, split into two stellar mass bins. This figure complements Figure \ref{fig:deltla_lblt} in the main text, which shows only the median trends; the violin distributions shown here provide the full distributional context underlying those medians.

\section{Morphological Fraction Trends} \label{appendix:morph_fractions}

\begin{figure*}[htb]
    \centering
    \includegraphics[width=1\linewidth]{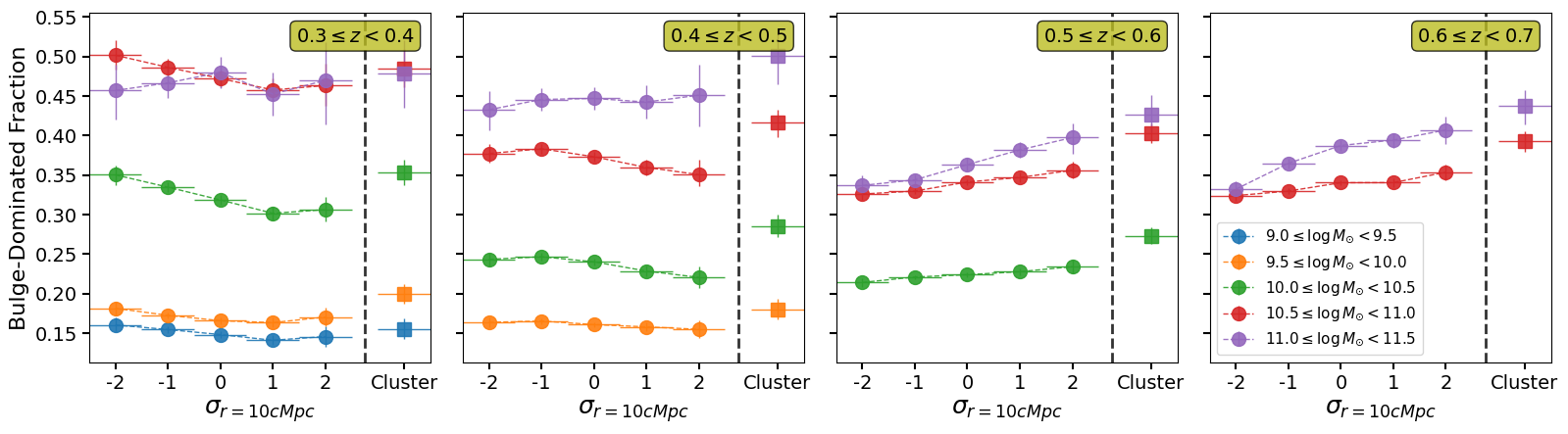}
    \caption{Fraction of bulge-dominated ($L_B/L_T \geq 0.6$) galaxies across the six environmental bins. Error bars represent uncertainties propagated from the $L_B/L_T$ probability distribution functions for individual galaxies in each bin. Colors denote the six stellar-mass bins defined in the legend. In each panel, we only show mass bins that are above the stellar mass completeness threshold. The complementary disk-dominated fraction ($L_B/L_T \leq 0.4$) is not shown as it is nearly the mirror image of the trends shown here.}
    \label{fig:fracden}
\end{figure*}

As a complementary diagnostic to the \deltalblt{} analysis in \S\ref{subsec:lblt_v_env}, we recast the relation between structural parameter and environment in terms of fractions of morphological classes. This view follows the long-standing practice of expressing the morphology--environment correlation via morphological fractions but now evaluated at fixed stellar mass. Whereas \S\ref{subsec:lblt_v_env} exploited the continuous nature of $L_B/L_T$ and its full posteriors to robustly test for monotonic trends, the fraction-based approach compresses the information into easily interpretable population ratios. Presenting the results in this form also enables direct comparison with many previous studies, which have typically reported trends in terms of disk- and bulge-dominated fractions. Such comparisons should, however, be made with care, since our classifications are based on structural parameter measurements and may not correspond exactly to visually defined morphology classes (e.g., a disk-dominated light profile does not necessarily correspond to a visually classified spiral galaxy).

In each redshift slice, we divide the sample into five stellar-mass bins and measure, for every environmental bin, the fraction of disk- and bulge-dominated galaxies. We show the bulge-dominated fractions in Figure \ref{fig:fracden}. The colors of the lines and points in Figure \ref{fig:fracden} correspond to the stellar-mass bins shown in the legend, and only bins above the stellar-mass completeness limit are included. Error bars represent uncertainties propagated from the $L_B/L_T$ probability distribution functions for each galaxy and are derived using the same PDF-sampling procedure described in \S \ref{subsec:lblt_v_env}: For each galaxy, we draw 5000 realizations from its $L_B/L_T$ distribution and calculate disk- and bulge-dominated fractions for each bootstrap realization. The plotted points mark the median of the fraction distribution, and the error bars span the 16th - 84th percentile range (i.e., $1\sigma$).

At $z < 0.5$, Figure \ref{fig:fracden} shows that stellar mass is the dominant driver of morphology, with the fraction of bulge-dominated galaxies systematically increasing with stellar mass in each redshift slice. Within individual mass bins, the bulge-dominated fractions show no evidence for a systematic monotonic dependence on environment. The cluster points, however, provide a useful contrast: in most cases, the bulge-dominated fractions in clusters lie systematically above the field sequence. This offset suggests that cluster-specific processes introduce a mild bias toward bulge-dominated systems even at a fixed stellar mass, consistent with the quantitative tests presented in \S\ref{subsec:lblt_v_env}.

In the two higher redshift slices, Figure \ref{fig:fracden} reveals more consistent monotonic trends across all six environmental bins. The fraction of bulge-dominated galaxies rises steadily as one moves from underdense to overdense environments. Note that this environmental trend is still secondary when compared to the significant separation between the fractions among individual mass bins. This aligns well with the quantitative results of \S \ref{subsec:lblt_v_env}, where weak but statistically significant correlations were identified at $z \geq 0.5$ for $\log \mathrm{M}/\mathrm{M}_\odot \geq 10.25$. These trends suggest that at these cosmic epochs, environmental processes capable of transforming morphology are at play at both large and cluster scales.

Overall, Figure~\ref{fig:fracden} provides a confirmation of the results established in \S\ref{subsec:lblt_v_env}: stellar mass is the primary determinant of morphology, but denser environments exert a measurable secondary influence.

\begin{acknowledgments}
C.I., A.G., and A.J.C. acknowledge support from the DiRAC Institute in the Department of Astronomy at the University of Washington. The DiRAC Institute is supported through generous gifts from the Charles and Lisa Simonyi Fund for Arts and Sciences and the Washington Research Foundation.

A.G. is supported by an LSST-DA Catalyst Fellowship; this publication was thus made possible through the support of Grant 62192 from the John Templeton Foundation to LSST-DA. The opinions expressed in this publication are those of the author(s) and do not necessarily reflect the views of LSST-DA or the John Templeton Foundation. L.O.V.E. is also partially supported by 62192.

A.G. acknowledges support from the University of Washington eScience Institute for the UW Data Science Postdoctoral Fellowship.

C.M.U. acknowledges support from the National Science Foundation under Grant No. AST-2407751.

This material is based in part upon work while L.O.V.E. served at the National Science Foundation. Any opinions, findings, and conclusions or recommendations are solely those of the authors and do not necessarily reflect the views of the National Science Foundation or the Federal government.

B.E.R. acknowledges support from the National Science Foundation under Grant No. AST-2307158.

The Hyper Suprime-Cam (HSC) collaboration includes the astronomical communities of Japan and Taiwan, and Princeton University. The HSC instrumentation and software were developed by the National Astronomical Observatory of Japan (NAOJ), the Kavli Institute for the Physics and Mathematics of the Universe (Kavli IPMU), the University of Tokyo, the High Energy Accelerator Research Organization (KEK), the Academia Sinica Institute for Astronomy and Astrophysics in Taiwan (ASIAA), and Princeton University. Funding was contributed by the FIRST program from Japanese Cabinet Office, the Ministry of Education, Culture, Sports, Science and Technology (MEXT), the Japan Society for the Promotion of Science (JSPS), Japan Science and Technology Agency (JST), the Toray Science Foundation, NAOJ, Kavli IPMU, KEK, ASIAA, and Princeton University. 

This paper makes use of software developed for the Large Synoptic Survey Telescope. We thank the LSST Project for making their code available as free software at \href{http://dm.lsst.org}{http://dm.lsst.org}.

Based, in part, on data collected at the Subaru Telescope and retrieved from the HSC data archive system, which is operated by Subaru Telescope and Astronomy Data Center at National Astronomical Observatory of Japan.
\end{acknowledgments}

\software{{Numpy \citep{Harris2020ArrayNumPy},
          Scipy \citep{scipy},
          Astropy \citep{2013A&A...558A..33A, 2018AJ....156..123A},
          Pandas \citep{pandas},
          Matplotlib \citep{Hunter2007Matplotlib:Environment}
          }}

\clearpage
\bibliography{References}
\bibliographystyle{aasjournalv7}

\end{CJK*}
\end{document}